\RequirePackage{fix-cm}
\documentclass[twocolumn]{svjour3}          % twocolumn
\smartqed  % flush right qed marks, e.g. at end of proof
\usepackage[utf8]{inputenc} %para colocar os acentos PT
\usepackage{graphicx}
\usepackage{epsfig}
\usepackage{subfigure}
\usepackage{calc}
\usepackage{amssymb}
\usepackage{amstext}
\usepackage{amsmath}

\usepackage{amsthm}
\usepackage{multicol}
\usepackage{pslatex}
\usepackage{longtable}
\usepackage[inline]{enumitem}
\usepackage{caption,multicol,booktabs,array} % Tabelas 
\usepackage[numbers]{natbib}
\usepackage{microtype}
\usepackage{verbatim} 
\usepackage{comment}
\usepackage[table,xcdraw]{xcolor}
\usepackage{diagbox}
\begin{document}

\title{Code smells detection and visualization
}
\subtitle{A systematic literature review}


%\begin{comment}
\author{First Author         \and
        Second Author %etc.
}

\author{José Pereira dos Reis
\and Fernando Brito e Abreu
\and Glauco de Figueiredo Carneiro
\and Craig Anslow
}


\institute{José Pereira dos Reis \at
              Iscte - Instituto Universitário de Lisboa, ISTAR-Iscte, Lisboa, Portugal\\
              \email{jvprs@iscte-iul.pt}           %  \\
           \and
            Fernando Brito e Abreu \at
              Iscte - Instituto Universitário de Lisboa, ISTAR-Iscte, Lisboa, Portugal\\
              \email{fba@iscte-iul.pt}
             \and
            Glauco de Figueiredo Carneiro \at
              Universidade Salvador (UNIFACS), Salvador, Bahia, Brazil\\
              \email{glauco.carneiro@unifacs.br}
             \and
            Craig Anslow \at
             Victoria University of Wellington, Kelburn, New Zealand\\
            \email{craig@ecs.vuw.ac.nz}
}

%\end{comment}

\date{Received: date / Accepted: date}

\maketitle

\begin{abstract}
\vspace{0pt}
\textit{}\textit{Context}: Code smells tend to compromise software quality and also demand more effort by developers to maintain and evolve the application throughout its life-cycle. They have long been catalogued with corresponding mitigating solutions called refactoring operations. Researchers have argued that due to the subjectiveness of the code smells detection process, proposing an effective use of automatic support for this end is a non trivial task.

%However, despite the successful initiatives of integrating many of the latter in current IDEs (e.g., Eclipse), code smells detection has not gained the same status.%

\textit{Objective}: This Systematic Literature Review (SLR) has a twofold goal: the first is to identify the main code smells detection techniques and tools discussed in the literature, and the second is to analyze to which extent visual techniques have been applied to support the former.
 
\textit{Method}: Over eighty primary studies indexed in major scientific repositories were identified by our search string in this SLR. Then, following existing best practices for secondary studies, we applied inclusion/exclusion criteria to select the most relevant works, extract their features and classify them.

\textit{Results}: We found that the most commonly used approaches to code smells detection are search-based (30.1\%), metric-based (24.1\%), and symptom-based approaches (19.3\%). Most of the studies (83.1\%) use open-source software, with the Java language occupying the first position (77.1\%). In terms of code smells, \textit{God Class} (51.8\%), \textit{Feature Envy} (33.7\%), and \textit{Long Method} (26.5\%) are the most covered ones. Machine learning (ML) techniques are used in 35\% of the studies, with genetic programming, decision tree, support vector machines (SVM) and association rules being the most used algorithms. Around 80\% of the studies only detect code smells, without providing visualization techniques. In visualization-based approaches several methods are used, such as: city metaphors, 3D visualization techniques, interactive ambient visualization, polymetric views, or graph models.
 
\textit{Conclusions}: This paper presents an up-to-date review on the state-of-the-art techniques and tools used for code smells detection and visualization. We confirm that the detection of code smells is a non trivial task, and there is still a lot of work to be done in terms of: reducing the subjectivity associated with the definition and detection of code smells; increasing the diversity of detected code smells and of supported programming languages; constructing and sharing oracles and datasets to facilitate the replication of code smells detection and visualization techniques validation experiments.
 
\keywords{systematic literature review \and code smells \and code smells detection \and code smells visualization \and software quality \and software maintenance}

\end{abstract}


%\linenumbers

\section{Introduction}
\label{sec:introduction}
Software maintenance has historically been the Achilles' heel of the software life cycle \cite{abreu95b}. Maintenance tasks can be seen as incremental modifications to a software system that aim to add or adjust some functionality or to correct some design flaws and fix some bugs. It has been found that feature addition, modification, bug fixing, and design improvement can cost as much as 80\% of the total software development cost \cite{Travassos:1999:DDO:320384.320389}. Code smells (CS), also called ``bad smells'', are associated with symptoms of software maintainability problems \cite{Yamashita2013d}. They often correspond to the violation of fundamental software design principles and negatively impact its future quality. Those weaknesses in design may slow down software evolution (e.g. due to code misunderstanding) or increase the risk of bugs or failures in the future. In this context, the detection of CS or anti-patterns (undesirable patterns, said to be recipes for disaster \cite{Brown1998}) is a topic of special interest, since it prevents code misunderstanding and mitigates potential maintenance difficulties. According to the authors of \cite{Singh2017}, there is a subtle difference between a CS and an anti-pattern: the former is a kind of warning for the presence of the latter. Nevertheless, in the remaining paper, we will not explore that slight difference and only refer to the CS concept.

%The anti-patterns are higher-level design defects (for example, \textit{God Class/Blob}, a class which has too many functions, so it violates high cohesion and probably low coupling, too), CS are lower-level defects, also called bad smells (such as \textit{Long Method}, or \textit{Duplicated Code}).

Code smells have been catalogued. The most widely used catalog was compiled by Martin Fowler \cite{Fowler1999}, and describes 22 CS. Other researchers, such as van Emden and Moonen \cite{Emden2002}, have subsequently proposed more CS. In recent years, CS have been cataloged for other object-oriented programming languages, such as Matlab \cite{Gerlitz2015DetectionAH}, Python \cite{Chen2016} and Java Android-specific CS \cite{Palomba2017Android, Kessentini2017Android}, which confirms the increasing recognition of their importance.

Manual CS detection requires code inspection and human judgment, and is therefore unfeasible for large software systems. Furthermore, CS detection is influenced (and hampered) by the subjectivity of their definition, as reported by Mantyla et al \cite{Mantyla2004}, based on the results of experimental studies. They observed the highest inter-rater agreements between evaluators for simple CS, but when the subjects were asked to identify more complex CS, such as \textit{Feature Envy}, they had the lowest coefficient of concordance. The main reason reported for this fact was that participants had no clear idea of what the \textit{Feature Envy} CS was. In other words, they suggested that experience may mitigate the subjectivity issue and indeed they observed that experienced developers reported more complex CS than novices did. However, they also concluded that the CS' commonsense detection rules expressed in natural language can also cause misinterpretation.

Automated CS detection, mainly in object-oriented systems, involves the use of source code analysis techniques, often metrics-based \cite{Lanza2006}. Despite research efforts dedicated to this topic in recent years, the availability of automatic detection tools for practitioners is still scarce, especially when compared to the number of existing detection methods (see section \ref{subsec:Evaluationtechniques}). 

Many researchers proposed CS detection techniques. However, most studies are only targeted to a small range of existent CS, namely \textit{God Class}, \textit{Long Method} and \textit{Feature Envy}. Moreover, only a few studies are related with the application of calibration techniques in CS detection (see section \ref{subsec:Codesmellsdetection} and \ref{subsec:Evaluationtechniques}).

Considering the diversity of existing techniques for CS detection, it is important to group the different approaches into categories for a better understanding of the type of technique used. Thus, we will classify the existing approaches into seven broad categories, according to the classification proposed by Kessentini et al. \cite{Kessentini2014}: manual approaches, symptom-based approaches, metric-based approaches, probabilistic approaches, visualization-based approaches, search-based approaches and cooperative-based approaches.

A factor that exacerbates the complexity of CS detection is that practitioners have to reason at different abstraction levels: some CS are found at the class level, others at the method level and even others encompass both method and class levels simultaneously (e.g. \textit{Feature Envy}). This means that once a CS is detected, its extension / impact must be conveyed to the developer, to allow him to take appropriate action (e.g. a refactoring operation). For instance, the representation of a \textit{Long Method} (circumvented to a single method) will be rather different from that of a \textit{Shotgun Surgery} that can spread across a myriad of classes and methods. Therefore, besides the availability of appropriate CS detectors, we need suggestive and customized CS visualization features, to help practitioners understand their manifestation. Nevertheless, there are only a few primary studies aimed at CS visualization.

We classify CS visualization techniques in two categories: (i) the detection is done through a non-visual approach, the visualization being performed to show CS location in the code itself, (ii) the detection is performed through a visual approach. In this Systematic Literature Review (SLR) we approach those two categories.

Most of the proposed CS visualization techniques show them inside the code itself. This approach works for some systems, but when we are in the presence of large legacy systems, it is too detailed for a global refactoring strategy. Thus, a more macro approach is required, without losing detail, to present CS in a more aggregated form. There are few primary studies in that direction.

Summing up, the main objectives for this review are:
\begin{itemize}
\item What are the main techniques for the detection of CS and their respective effectiveness reported in the literature?
\item What are the visual approaches and techniques reported in the literature to represent CS and therefore support practitioners to identify their manifestation?
\end{itemize}

The rest of this paper is organized as follows. Section \ref{sec:Relatedwork} describes the differences between this SLR and related ones. The subsequent section outlines the adopted research methodology (section \ref{sec:ResearchMethodology}). Then, SLR results and corresponding analyses are presented (section \ref{sec:ResultandAnalysis}). The answers to the Research Questions (RQ) are discussed in section \ref{sec:discussion}, and the concluding remarks, as well as scope for future research, are presented in section \ref{sec:conclusion}.

\section{Related work}
\label{sec:Relatedwork}

We will present the related work in chronological order.

Zhang et al. \cite{Zhang2010} presented a systematic review on CS, where more than 300 papers published from 2000 to 2009 in leading journals from IEEE, ACM, Springer and other publishers were investigated. After applying the selection criteria, the 39 most relevant ones were analyzed in detail. Different research parameters were investigated and presented from different perspectives. The authors revealed that \textit{Duplicated Code} is the most widely studied CS. Their results suggest that only a few empirical studies have been conducted to examine the impact of CS and therefore a phenomenon that was far from being fully understood.

Rattan et al. \cite{Rattan2013} performed a vast literature review to study software clones (aka \textit{Duplicate Code}) in general and software clone detection in particular. The study was based on a comprehensive set of 213 articles from a total of 2039 articles published in 11 leading journals and 37 premier conferences and workshops. An empirical evaluation of clone detection tools/techniques is presented. Clone management, its benefits and cross cutting nature is reported. A number of studies pertaining to nine different types of clones is reported, as well as thirteen intermediate representations and 24 match detection techniques. In conclusion, the authors call for an increased awareness of the potential benefits of software clone management and identify the need to develop semantic and model clone detection techniques. 

Rasool and Arshad \cite{Rasool2015} presented a review on several detection techniques and tools for mining CS. They classify selected CS detection techniques and tools based on their detection methods and analyze the results of the selected techniques. This study presented a critical analysis, where the limitations for the different tools are identified. The authors concluded, for example, that there is still no consensus on the common definitions of CS by the research community and there is a lack of standard benchmark systems for evaluating existing techniques.

Al Dallal \cite{AlDallal2015} performed a SLR on the possibilities of performing refactoring in object-oriented systems. The primary focus is on the detection of CS and covered 45 primary studies. Various approaches for the detection of CS were brought into limelight. The work revealed the open source systems potentially used by the researchers, and the author found that, among those systems, JHotDraw is the most used by researchers to validate their results. Similarly, the Java language was found to be the most reported language on refactoring studies.

Fernandes et al. \cite{Fernandes2016} presented the findings of a SLR on CS detection tools. They found in the literature a mention to 84 tools, but only 29 of them were available online for download. Altogether, these tools aim to detect 61 CS, by relying on at least six different detection techniques. The review results show that Java, C, and C++ are the top-three most covered programming languages for CS detection. The authors also present a comparative study of four detection tools with respect to two CS: \textit{Large Class} and \textit{Long Method}. Their findings support that tools provide redundant detection results for the same CS. Finally, this SLR concluded that \textit{Duplicated Code}, \textit{Large Class}, and \textit{Long Method} are the top-three CS that tools aim to detect.

Singh and Kaur \cite{Singh2017} published a SLR on refactoring with respect to CS. Although the title appears to focus on refactoring, different types of techniques for identifying CS and antipatterns are discussed in depth. The authors claim that this work is an extension of the one published in \cite{AlDallal2015}. They found 1053 papers in the first round, which they refined to 325 papers based on the title of the paper. Then, based on the abstract, they trimmed down that number to 267. Finally, a set of 238 papers was selected after applying inclusion and exclusion criteria. This SLR includes primary studies from the early ages of digital libraries till September 2015. Some conclusions regarding detection approaches were that 28.15\% of researchers applied automated detection approaches to discover the CS, while empirical studies are used by a total of 26.89\% of researchers . The authors also pointed out that Apache Xerces, JFreeChart and ArgoUML are among the most targeted systems that, for obvious reasons, are usually open source. They also reckon that \textit{God Class} and \textit{Feature Envy} are the most recurrently detected CS.

Gupta et al. \cite{Gupta2017} performed a SLR based on publications from 1999 to 2016 and 60 papers, screened out of 854, are deeply analyzed. The objectives of this SLR were to provide an extensive overview of existing research in the field of CS, identify the detection techniques and find out which are the CS that deserve more attention in detection approaches. This SLR identified that the \textit{Duplicate Code} CS receives most research attention and that very few papers report on the impact of CS. The authors conclude that most papers were focused on the detection techniques and tools and a significant correlation between detection techniques and CS has been performed on the basis of CS. They also identified four CS from Fowler's catalog, whose detection is not reported in the literature: \textit{Primitive Obsession}, \textit{Inappropriate Intimacy}, \textit{Incomplete Library Class} and \textit{Comments}.

Alkharabsheh et al. \cite{Alkharabsheh2018} performed a systematic mapping study where they analyzed 18 years of research into Design Smell Detection based on a comprehensive set of 395 articles published in different proceedings, journals, and book chapters. Some key findings for future trends include the fact that all automatic detection tools described in the literature identify Design Smells as a binary decision (having the smell or not), lack of human experts and benchmark validation processes, as well as demonstrating that Design Smell Detection positively influences quality attributes. The authors found an important problem which is the absence of an extensive Smell Corpus Design available in common to several detection tools.

Santos et al. \cite{SANTOS2018} investigating how CS impact the software development, the CS effect. They reached three main results: that the CS concept does not support the evaluation of quality design in practice activities of software development, i.e., there is still a lack of understanding of the effects of CS on software development; there is no strong evidence correlating CS and some important software development attributes, such as maintenance effort; and the studies point out that human agreement on CS detection is low. The authors suggest that to improve analysis on the subject, the area needs to better outline: (i) factors affecting human evaluation of CS; and (ii) a classification of types of CS, grouping them according to relevant characteristics.

Sabir et al. \cite{Sabir2018} investigating the key techniques employed to identify smells in different paradigms of software engineering from object-oriented (OO) to service-oriented (SO). They performed a SLR based on publications from January 2000 to December 2017 and selected 78 papers. The authors concluded that: the most used CS in the literature are \textit{Feature Envy}, \textit{God Class}, \textit{Blob}, and \textit{Data Class}; Smells like the yo-yo problem, unnamed coupling, intensive coupling, and interface bloat received considerably less attention in the literature; Mainly two techniques in the detection of smells are used in the literature static source code analysis and dynamic source code analysis based on dynamic threshold adaptation, e.g., using a genetic algorithm, instead of fixed thresholds for smell detection. 

The SLR proposed by Azeem et al. \cite{AZEEM2019} investigated the usage of ML approaches in the field of CS between 2000 and 2017. From an initial set of 2456 papers, they found that 15 papers actually adopted ML approaches. They studied them from four different perspectives: (i) CS considered, (ii) setup of ML approaches, (iii) design of the evaluation strategies, and (iv) a meta-analysis on the performance achieved by the models proposed so far. The authors concluded that: the most used CS in the literature are \textit{God Class}, \textit{Long Method}, \textit{Functional Decomposition}, and \textit{Spaghetti Code};  Decision Trees and Support Vector Machines are the most commonly used ML algorithms for CS detection; several open issues and challenges exist that the research community should focus on in the future. Finally, they argue that there is still room for the improvement of ML techniques in the context of CS detection.

Kaur \cite{Kaur2019} examined 74 primary studies covering the impact of CS on software quality attributes. The results indicate that the impact of CS on software quality is not uniform as different CS have the opposite effect on different software quality attributes. The author observed that most empirical studies reported the incoherent impact of CS on quality. This contradictory impact may be due to the size of the data set considered or the programming language in which the data sets are implemented. Thus, Kaur concludes the actual impact of CS on software quality is still unclear and needs more attention.

The scope and coverage of the current SLR goes beyond those of aforementioned SLRs, mainly because it also covers the CS visualization aspects. The latter is important to show programmers the scope of detected CSs, so that they decide whether they want to proceed to refactoring or not. A good visualization becomes even more important if one takes into account the subjectivity existing in the definition of CS, which leads to the detection of many false positives.

\section{Research Methodology}
\label{sec:ResearchMethodology}
In contrast to a non-structured review process, a SLR reduces bias and follows a precise and rigorous sequence of methodological steps to review research literature \cite{brereton2007lessons} and \cite{KitchenhametAl.2007}. SLRs rely on well-defined and evaluated review protocols to extract, analyze, and document results, as the stages conveyed in Figure \ref{fig:SLR_stages}. This section describes the methodology applied for the phases of planning, conducting and reporting the review.

\subsection{Planning the Review}
\label{subsec:PlanningReview}

\noindent {\bf Identify the needs for a systematic review.} Search for evidences in the literature regarding the main techniques for CS detection and visualization, in terms of (i) strategies to detect CS, (ii) effectiveness of CS detection techniques, (iii) approaches and techniques for CS visualization.

\noindent \textbf{The Research Questions.} We aim to answer the following questions, by conducting a methodological review of existing research:

\noindent \textbf{RQ1}. \textit{Which techniques have been reported in the literature for the detection of CS?} The list of the main techniques reported in the literature for the detection of CS can provide a comprehensive view for both practitioners and researchers, supporting them to select a technique that best fit their daily activities, as well as highlighting which of them deserve more effort to be analyzed in future experimental studies.

\noindent \textbf{RQ2}. \textit{What literature has reported on the effectiveness of techniques aiming at detecting CS?} The goal is to compare the techniques among themselves, using parameters such as accuracy, precision and recall, as well as the classification of automatic, semi-automatic or manual.

\noindent \textbf{RQ3}. \textit{What are the approaches and resources used to visualize CS and therefore support the practitioners to identify CS occurrences?} The visualization of CS occurrences is a key issue for its adoption in the industry, due to the variety of CS, possibilities of location within code (e.g. in methods, classes, among classes), and dimension of the code for a correct identification of the CS.

These three research questions are somehow related to each other. In fact, any detection algorithm after being implemented, should be tested and evaluated to verify its effectiveness, which causes RQ1 and RQ2 to be closely related. RQ3 encompasses two possible situations: i) CS detection is done through visual techniques, and ii) visual approaches are only used for representing CS previously detected with other techniques; therefore, there is also a close relationship between RQ1 and RQ3. 

\noindent \textbf{Publications Time Frame.} We conducted a SLR in journals, conferences papers and book chapters from January 2000 to June 2019.  
 
\subsection{Conducting the Review}
\label{subsec:ConductingReview}
This phase is responsible for executing the review protocol. 

\noindent \textbf{Identification of research.} Based on the research questions, keywords were extracted and used to search the primary study sources. The search string is presented as follows and used the same strategy cited in \cite{chen2011systematic}:

\begin{quote}
\textit{(``code smell" OR ``bad smell") AND (visualization OR visual OR representation OR identification OR detection) AND (methodology OR approach OR technique OR tool)}
\end{quote}

\noindent \textbf{Selection of primary studies.} The following steps guided the selection of primary studies. 

\textit{Stage 1 - Search string results automatically obtained from the engines} - Submission of the search string to the following repositories: ACM Digital Library, IEEE Xplore, ISI Web of Science, Science Direct, Scopus and Springer Link. The justification for the selection of these libraries is their relevance as sources in software engineering \cite{zhang2011identifying}. The search was performed using the specific syntax of each database, considering only the title, keywords, and abstract. The search was configured in each repository to select only papers carried out within the prescribed period. The automatic search was complemented by a backward snowballing manual search, following the guidelines of Wohlin \cite{Wohlin2014}. The duplicates were discarded.

\textit{Stage 2 - Read titles \& abstracts to identify potentially relevant studies} - Identification of potentially relevant studies, based on the analysis of title and abstract, discarding studies that are clearly irrelevant to the search. If there was any doubt about whether a study should be included or not, it was included for consideration on a later stage.

\textit{Stage 3 - Apply inclusion and exclusion criteria on reading the introduction, methods and conclusion} - Selected studies in previous stages were reviewed, by reading the introduction, methodology section and conclusion. Afterwards, inclusion and exclusion criteria were applied (see Table \ref{table:TableInclusionCriteria} and Table \ref{table:TableExclusionCriteria}). At this stage, in case of doubt preventing a conclusion, the study was read in its entirety. 

\begin{table}[htbp]
\caption{Inclusion criteria}
\label{table:TableInclusionCriteria} \centering
\smallskip
\begin{tabular}{|c|p{5cm}|}
 \hline
 \bf{ Criterion} & \bf{Description}\\
 \hline
 IC1 & The publication venue should be a “journal” or “conference proceedings” or "book".\\
 \hline
 IC2 & The primary study should be written in English.\\
 \hline
 IC3 & The primary work is an empirical study or have "lessons learned" (experience report).\\
 \hline
 IC4 & If several papers report the same study, the latest one will be included.\\
 \hline
 IC5 & The primary work addresses at least one of the research questions.\\
 \hline
\end{tabular}
\end{table}

\begin{table}[htbp]
\caption{Exclusion criteria}
\label{table:TableExclusionCriteria} \centering
\begin{tabular}{|c|p{5cm}|}
\hline
\bf{Criterion} & \bf{Description}\\
\hline
EC1 & Studies not focused on code smells.\\
\hline
EC2 & Short paper (less than 2000 words, excluding numbers) or unavailable in full text.\\
\hline
EC3 & Secondary and tertiary studies, editorials/prefaces, readers’ letters, panels, and poster-based short papers.\\
\hline
EC4 & Works published outside the selected time frame.\\
\hline
EC5 & Code Smells detected in non-object oriented programming languages.\\
\hline
\end{tabular}
\end{table}

The  reliability  of  the  inclusion and exclusion criteria of a publication in the SLR was assessed  by  applying  Fleiss’  Kappa \cite{Fleiss2013}.  Fleiss’  Kappa  is  a statistical  measure  for  assessing  the  reliability  of  agreement between a fixed number of raters when classifying items. We used the Kappa statistic \cite{McHugh2012} to measure the level of agreement between the researchers. Kappa result is based on the number of answers with the same result for both observers \cite{Landis1977}. Its maximum value is 1, when the researchers have almost perfect agreement, and it tends to zero or less when there are no agreement between them (Kappa can range from {-1} to +1). The higher the value of Kappa, the stronger the agreement. Table \ref{table:TableKappaResults} shows the interpretation of this coefficient according to Landis \& Koch \cite{Landis1977}.

\begin{table}[htbp]
\caption{Interpretation of the Kappa results}
\label{table:TableKappaResults} \centering
\smallskip
\begin{tabular}{|c|l|}
 \hline
 \textbf{Kappa values} & \textbf{Degree of agreement}\\
 \hline
 \textless 0.00 & Poor\\
 \hline
 0.00 - 0.20 & Slight\\
 \hline
 0.21 - 0.40 & Fair(Weak)\\
 \hline
 0.41 - 0.60 & Moderate\\
 \hline
 0.61 - 0.80 & Substantial (Good)\\
 \hline
 0.81 - 1.00 & Almost perfect (Very Good)\\
 \hline
\end{tabular}
\end{table}

We asked two seniors researchers to classify, individually, a sample of 31 publications to analyze the degree of agreement in the  selection process through the Fleiss’ Kappa \cite{Fleiss2013}.  The selected sample was the set of the most recent publications (last 2 years) from phase 2. The result of the degree of agreement showed a substantial level of agreement between the two researchers (Kappa = 0.653).

The 102 studies resulting from this phase are listed in Appendix \ref{AppendixB}.

\textit{Stage 4 - Obtain primary studies and make a critical assessment of them} - A list of primary studies was obtained and later subjected to critical examination using the 8 quality criteria set out in Table \ref{table:TableQualityCriteria}. Some of these quality criteria were adapted from those proposed by Dyba and Dings{\o}yr \cite{Dyba2008}. In the QC1 criterion we evaluated venue quality based on its presence in the CORE rankings portal\footnote{http://www.core.edu.au/}. In the QC4 criterion, the relevance of the study to the community was evaluated based on the citations present in Google Scholar\footnote{https://scholar.google.com/} using the method of Belikov and Belikov \cite{Belikov2015}. The grading of each of the 8 criteria was done on a dichotomous scale ("YES"=1 or "NO"=0). For each selected primary study, its quality score was computed by summing up the scores of the answers to all the 8 questions. A given paper satisfies the Quality Assessment criteria if reaches a rating higher (or equal) to 4. Among the 102 papers resulting from stage 3, 19 studies [11, 16, 19, 22, 32, 36, 57, 71, 73, 77, 82, 83, 85, 86, 87, 90, 91, 95, 102] (see Appendix \ref{AppendixB}) were excluded because they did not reach the minimum score of 4 (Table \ref{table:ResultsQualityCriteria}), while 83 passed the Quality Assessment criteria. All 83 selected studies are listed in Appendix \ref{AppendixA} and the details of the application of the quality assessment criteria are presented in Appendix \ref{AppendixC}.

\begin{table}[h]
\caption{Quality criteria}
\label{table:TableQualityCriteria} \centering
\smallskip
\begin{tabular}{|c|p{6cm}|}
 \hline
 \textbf{Criterion} & \textbf{Description}\\
 \hline
 QC1 & Is the venue recognized in CORE rankings portal?\\
 \hline
 QC2 & Was the data collected in a way that addressed the research issue?\\
 \hline
 QC3 & Is there a clear statement of findings?\\
 \hline
 QC4 &  Is the relevance for research or practice recognized by the community?\\
 \hline
 QC5 & Is there an adequate description of the validation strategy?\\
 \hline
 QC6 & The study contains the required elements to allow replication?\\
 \hline
 QC7 & The evaluation strategies and metrics used are explicitly reported?\\
 \hline
 QC8 & Is a CS visualization technique clearly defined?\\
 \hline
\end{tabular}
\end{table}

\begin{table}[h]
\caption{Number of studies by score obtained after application of the quality assessment criteria}
\label{table:ResultsQualityCriteria} \centering
\smallskip
\begin{tabular}{|c|c|c|}
 \hline
 \textbf{Resulting score} & \textbf{Number of studies} & \textbf{\% studies }\\
 \hline
 1 & 3 & 2.9\%\\
 \hline
 2 & 4 & 3.9\%\\
 \hline
 3 & 12 & 11.8\%\\
 \hline
 4 & 15 & 14.7\%\\
 \hline
 5 & 30 & 29.4\%\\
 \hline
 6 & 32 & 31.4\%\\
 \hline
 7 & 6 & 5.9\%\\
 \hline
 8 & 0 & 0.0\%\\
 \hline
\end{tabular}
\end{table}

\noindent {\bf Data extraction.} All relevant information on each study was recorded on a spreadsheet. This information was helpful to summarize the data and map it to its source. The following data were extracted from the studies: (i) name and authors; (ii) year; (iii) type of article (journal, conference, book chapter); (iv) name of conference, journal or book; (v) number of Google Scholar citations at the time of writing this paper; (vi) answers to research questions; (vii) answers to quality criteria. 

\noindent {\bf Data Synthesis.} This synthesis is aimed at grouping findings from the studies in order to: identify the main concepts about CS detection and visualization, conduct a comparative analysis on the characteristics of the study, type of method adopted, and issues regarding three research questions (\textit{RQ1, RQ2} and \textit{RQ3}) from each study. Other information was synthesized when necessary. We used the meta-ethnography method \cite{Noblit1988} as a reference for the process of data synthesis.

\begin{figure*}[!ht]
 \centering
 {\epsfig{file = 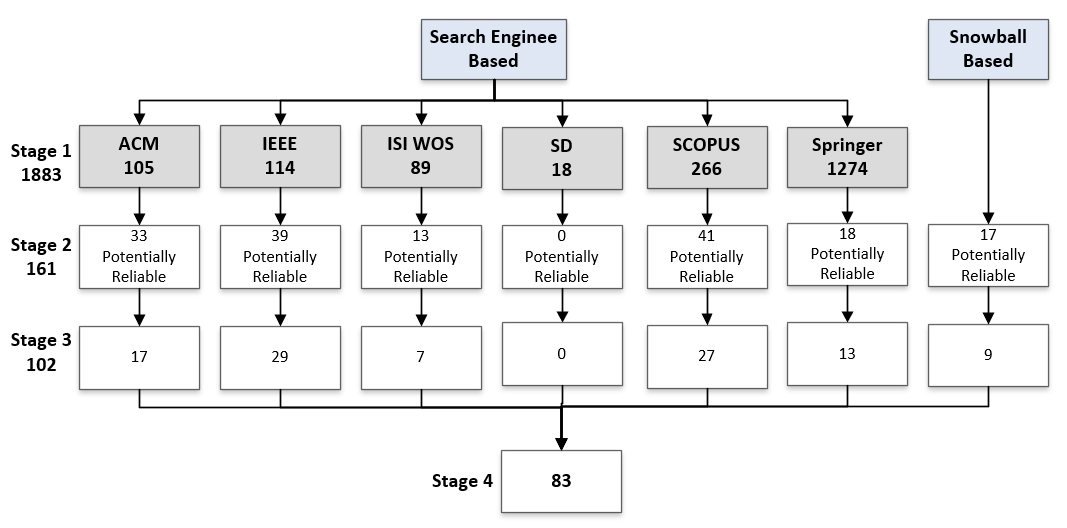, width = 14cm}}
 \caption{Stages of the study selection process}
 \label{fig:SLR_stages}
\end{figure*}

\noindent {\bf Conducting the Review.} We started the review with an automatic search followed by a manual search, to identify potentially relevant studies and afterwards applied the inclusion/exclusion criteria. We had to adapt the search string in some engines without losing its primary meaning and scope. The manual search consisted in studies published in conference proceedings, journals and books, that were included by the authors through backward snowballing in primary studies. These studies were equally analyzed regarding their titles and abstracts. Figure \ref{fig:SLR_stages} conveys them as 17 studies. We tabulated everything on a spreadsheet so as to facilitate the subsequent phase of identifying potentially relevant studies. Figure \ref{fig:SLR_stages} presents the results obtained from each electronic database used in the search, which resulted in 1866 articles considering all databases.

\noindent {\bf Potentially Relevant Studies.} The results obtained from both the automatic and manual search were included on a single spreadsheet. Papers with identical title, author(s), year and abstract were discarded as redundant. At this stage, we registered an overall of 1883 articles, namely 1866 from the automated search plus 17 from the separate manual search \textit{(Stage 1)}. We then read titles and abstracts to identify relevant studies resulting in 161 papers \textit{(Stage 2)}. At \textit{Stage 3} we read introduction, methodology and conclusion in each study and then we applied the inclusion and exclusion criteria, resulting in 102 papers. In \textit{Stage 4}, after applying the quality criteria \textit{(QC)} the remaining 83 papers were analysed to answer the three research questions - RQ1, RQ2 and RQ3.

\section{Results and Analysis}
\label{sec:ResultandAnalysis}
\begin{figure}[!ht]
 \centering
 {\epsfig{file = 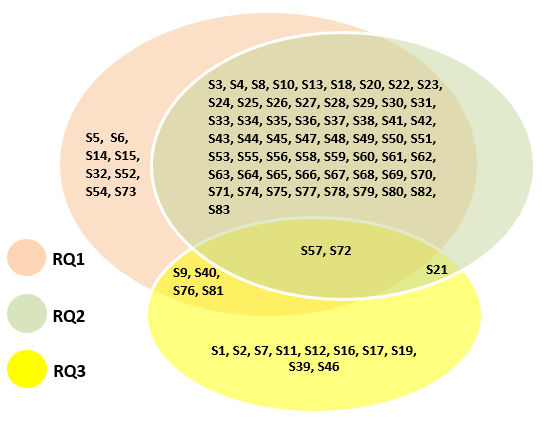, width = 8cm}}
 \caption{Selected studies per research question (RQ)}
 \label{fig:slrfinal}
\end{figure}

This section presents the results of this SLR to answer research questions RQ1, RQ2 and RQ3, based on the quality criteria and findings (F). In Figure \ref{fig:Findings-resume} we present a summary of the main findings. Figure \ref{fig:slrfinal} conveys the selected studies and the respective research questions they focus on. As can be seen in the same Figure, 72 studies addressed issues related to RQ1, while 61 studies discussed RQ2 issues and, finally, 17 papers addressed RQ3 issues. All selected studies are listed in Appendix \ref{AppendixA} and referenced as \textit{"S"} followed by the number of the paper. 

\subsection{Overview of studies}
\label{subsec:Overviewstudies}
 
\begin{figure}[!ht]
 \centering
 {\epsfig{file = 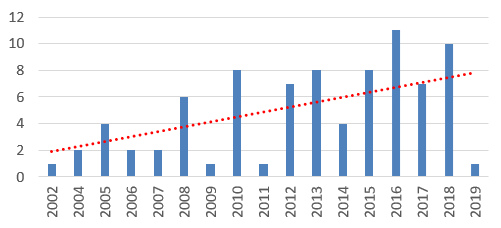, width = 8cm}}
 \caption{Trend of publication years}
 \label{fig:Pub-year}
\end{figure}

 The study selection process (Figure \ref{fig:SLR_stages}) resulted in 83 studies selected for data extraction and analysis. Figure \ref{fig:Pub-year} depicts the temporal distribution of primary studies. Note that 78.3\% primary studies have been published after 2009 (last 10 years) and that 2016 and 2018 were the years that had the largest number of studies published. This indicates that, although the human factor in software engineering has been acknowledged and researched since the 1970s, research focusing in CS detection is much more recent, with the vast majority of the studies developed in the last decade.
 
\begin{figure}[!ht]
 \centering
 {\epsfig{file = 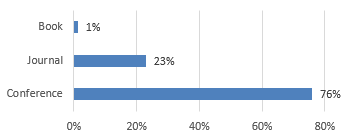, width = 8cm}}
 \caption{Type of publication venue}
 \label{fig:pub-venue}
\end{figure}

In relation to the type of publication venue (Figure \ref{fig:pub-venue}), the majority of the studies were published in conference proceedings 76\%, followed by journals with 23\%, and 1\% in books.

Table \ref{table:Tabletenmostcitedstudies} presents the top ten studies included in the review, according to Google Scholar citations in September 2019\footnote{Data obtained in 22/09/2019}. These studies are evidences of the relevance of the issues discussed in this SLR and the influence these studies exert on the literature, as can be confirmed by their respective citation numbers. Table \ref{table:Tabletenmostcitedstudies} shows an overview of the distribution of the most relevant studies according to the addressed research questions. In the following paragraphs, we briefly describe these studies, by decreasing order of impact.

\begin{table}[htpb]
\caption{Top-ten cited papers, according to Google Scholar}
\label{table:Tabletenmostcitedstudies} \centering
\smallskip
\begin{tabular}{|c|c|c|}
 \hline
 \textbf{Studies} & \textbf{Cited by} & \textbf{Research Question}\\
 \hline
 S9 & 964 & RQ1 and RQ3 \\
 \hline
 S26 & 577 & RQ1 and RQ2\\
 \hline
 S3 & 562 & RQ1 and RQ2 \\
 \hline
 S1 & 423 & RQ3 \\
 \hline
 S10 & 245 & RQ1 and RQ2 \\
 \hline
 S4 & 240 & RQ1 and RQ2 \\
 \hline
 S7 & 184 & RQ3 \\
 \hline
 S21 & 174 & RQ1 and RQ2 \\
 \hline
 S45 & 157 & RQ1 and RQ2 \\
 \hline
 S15 & 156 & RQ1 and RQ3 \\
 \hline
\end{tabular}
\end{table}

A brief review of each of the top cited paper follows:

\textbf{[S9]} - RQ1 and RQ3 are addressed in this paper that got the highest number of citations. It introduces a systematic way of detecting CS by defining detection strategies based in four steps: Step 1: Identify Symptoms; Step 2: Select Metrics; Step 3: Select Filters; Step 4: Compose the Detection Strategy. It describes how to explore quality metrics, set thresholds for these metrics, and create a set of rules to identify CS. Finally, visualization techniques are used to present the detection result, based in several metaphors. 

\textbf{[S26]} - The DECOR method for specifying and detecting code and design smells is introduced. This approach allows specifying smells at a high level of abstraction using a consistent vocabulary and domain-specific language for automatically generating detection algorithms. Four design smells are identified by DECOR, namely \textit{Blob}, \textit{Functional Decomposition}, \textit{Spaghetti Code}, and \textit{Swiss Army Knife}, and the algorithms are evaluated in terms of precision and recall. This study addresses RQ1 and RQ2, and is one of the most used studies for validation / comparison of results in terms of accuracy and recall of detection algorithms.

\textbf{[S3]} - Issues related to RQ1 and RQ2 are discussed. This paper proposes a mechanism called “detection strategies” for producing a metric-based rules approach to detect CS with detection strategies, implemented in the IPLASMA tool. This method captures deviations from good design principles and consists of defining a list of rules based on metrics and their thresholds for detecting CS.

\textbf{[S1]} - A visualization approach supported by the jCOSMO tool, a CS browser that performs fully automatic detection and visualizes smells in Java source code, is proposed. This study focuses its attention on two CS, related to Java programming language, i.e., \textit{instanceof} and \textit{typecast}. This paper discusses issues related to RQ3.

\textbf{[S10]} - This paper addresses RQ1 and RQ2 and proposes the Java Anomaly Detector (JADET) tool for detecting object usage anomalies in programs. JADET uses concept analysis to infer properties that are nearly always satisfied, and it reports the failures as anomalies. This approach is based on identifying usage patterns.

\textbf{[S4]} - This paper presents a metric-based heuristic detection technique able of identifying instances of two CS, namely \textit{Lazy Class} and \textit{Temporary Field}. It proposes a template to describe CS systematically, that consists of three main parts: a CS name, a text-based description of its characteristics, and heuristics for its detection. An empirical study is also reported, to justify the choice of metrics and thresholds for detecting CS. This paper discusses issues related to RQ1 and RQ2.

\textbf{[S7]} - This paper addresses RQ3 and presents a visualization framework for quality analysis and understanding of large-scale software systems. Programs are represented using metrics. The authors claim that their semi-automatic approach is a good compromise between fully automatic analysis techniques that can be efficient, but loose track of context, and pure human analysis that is slow and inaccurate.

\textbf{[S21]} - This paper proposes an approach based on Bayesian Belief Networks (BBN) to specify and detect CS in programs. Uncertainty is managed by BBN that implement the detection rules of DECOR [S26]. The detection outputs are probabilities that a class is an occurrence of a defect type. This paper discusses issues related to RQ1 and RQ2.

\textbf{[S45]} - This paper addresses RQ1 and RQ2, and proposes an approach called HIST (Historical Information for Smell deTection) to detect five different CS (\textit{Divergent Change}, \textit{Shotgun Surgery}, \textit{Parallel Inheritance}, \textit{Feature Envy}, and \textit{Blob}). HIST explores change history information mined from versioning systems to detect CS, by analyzing co-changes among source code artifacts over time.

\textbf{[S15]} - This last paper in the top ten most cited ones addresses RQ1 and RQ3. It presents an Eclipse plug-in (JDeodorant) that automatically identifies Type-Checking CS in Java source code, and allows their elimination by applying appropriate refactorings. JDeodorant is one of the most used tools for validation / comparison of results in terms of accuracy and recall of detection algorithms.

\subsection{Approach for CS detection (\textbf{F1})}
\label{subsec:ApproachCSdetection}
The first finding to be analyzed is the approach applied to detect CS, that is, the steps required to accomplish the detection process. For example, in the metric-based approach \cite{Lanza2006}, we need to know the set of source code metrics and corresponding thresholds for the target CS.

Considering the diversity of existing techniques for CS detection, it is important to group the different approaches into categories for a better understanding of the type of technique used. Thus, we will classify the existing approaches for CS detection into seven (7) broad categories, according to the classification presented by Kessentini et al. \cite{Kessentini2014}: metric-based approaches, search-based approaches, symptom-based approaches, visualization-based approaches, probabilistic approaches, cooperative-based approaches and manual approaches.

Classifying studies in one of the seven categories is not an easy task because some studies use intermediate techniques for their final technique. For example, several studies classified as symptom-based approaches use symptoms to describe CS, although detection is performed through a metric-based approach. 

Table \ref{table:CSdetectionapproaches} shows the classification of the studies in the seven broad categories. The most used approaches are search-based, metric-based, and symptom-based, being used in 30.1\%, 24.1\% and 19.3\% of the studies, respectively. The least used approaches are the cooperative-based and the manual ones, each being used in only one of the selected studies.

\begin{table}[htpb]
\caption{CS detection approaches used}
\label{table:CSdetectionapproaches} \centering
\smallskip
\begin{tabular}{|p{2cm}|p{1.1cm}|c|p{2cm}|}
 \hline
 \textbf{Approaches} & \textbf{Nº of studies} & \textbf{\% Studies} & \textbf{Studies}\\
 \hline
 Search-Based & 25 & 30.1\% & S5,S6,S14,S22, S24,S28,S33,S34, S36,S37,S42,S43, S45,S48,S51,S53, S55,S56,S71,S74, S75,S77,S78,S79, S83\\ \hline 
 Metric-Based & 20 & 24.1\% & S3,S9,S29,S38, S44,S47,S49,S52, S58,S60,S63,S64, S66,S67,S68,S70, S72,S73,S81,S82\\ \hline
 Symptom-based & 16 & 19.3\% & S4,S8,S13,S15, S20,S23,S26,S30, S31,S32,S52,S57, S59,S61,S62,S69\\ \hline
 Visualization-based & 12 & 14.5\% & S1,S2,S7,S11, S12,S16,S17,S19, S21,S39,S46,S76\\ \hline
 Probabilistic & 10 & 12.0\% & S10,S18,S25,S27, S35,S40,S50,S54, S65,S80\\ \hline
 Cooperative-based & 1 & 1.2\% & S41\\ \hline
 Manual & 1 & 1.2\% & S52\\ \hline
\end{tabular}
\end{table}

\subsubsection{Search-based approaches}
Search-based approaches are inspired by contributions in the domain of Search-Based Software Engineering (SBSE). SBSE uses search-based approaches to solve optimization problems in software engineering. Most techniques in this category apply ML algorithms. 
The major benefit of ML-based approaches is that they do not require great experts’ knowledge and interpretation. However, the success of these techniques depends on the quality of data sets to allow training ML algorithms.

\subsubsection{Metric-based approaches}
The metric-based approach is the most commonly used. The use of quality metrics to improve the quality of software systems is not a new idea and for more than a decade, metric-based CS detection techniques have been proposed. This approach consists in creating a rule, based on a set of metrics and respective thresholds, to detect each CS.

The main problem with this approach is that there is no consensus on the definition of CS, as such there is no consensus on the standard threshold values for the detection of CS. Finding the best fit threshold values for each metric is complicated because it requires a significant calibration effort \cite{Kessentini2014}. Threshold values are one of the main causes of the disparity in the results of different techniques.

\subsubsection{Symptom-based approaches}
To describe code-smell symptoms, different symptoms/notions are involved, such as class roles and structures. Symptom descriptions are later translated into detection algorithms. Kessentini et al. \cite{Kessentini2014} defines two main limitations to this approach:
\begin{itemize}
 \item there exists no consensus in defining symptoms;
 \item for an exhaustive list of CS, the number of possible CS to be manually described, characterized with rules and mapped to detection algorithms can be very large; as a consequence, symptoms-based approaches are considered as time-consuming and error-prone. 
\end{itemize}
Other authors \cite{Rasool2015} add more limitations, such as the analysis and interpretation effort required to select adequate threshold values when converting symptoms into detection rules. The precision of these techniques is low because of the different interpretations of the same symptoms.

\subsubsection{Visualization-based approaches}
Visualization-based techniques usually consist of a semi-automated process to support developers in the identification of CS. The data visually represented to this end is mainly enriched with metrics (metric-based approach) throughout specific visual metaphors.

This approach has the advantage of using visual metaphors, which reduces the complexity of dealing with a large amount of data. The disadvantages are those inherent to human intervention: (i) they require great human expertise, (ii) time-consuming, (iii) human effort, and (iv) error-prone. Thus, these techniques have scalability problems for large systems.

\subsubsection{Probabilistic approaches}
Probabilistic approaches consist essentially of determining a probability of an event, for example, the probability of a class being a CS. Some techniques consist on the use of BBN, considering the CS detection process as a fuzzy-logic problem or frequent pattern tree.

\subsubsection{Cooperative-based approaches}
Cooperative-based CS techniques are primarily aimed at improving accuracy and performance in CS detection. This is achieved by performing various activities cooperatively.

The only study that uses a cooperative approach is Boussaa et al. [S41]. According to the authors, the main idea is to evolve two populations simultaneously, where the first one generates a set of detection rules (combination of quality metrics) that maximizes the coverage of a base of CS examples and the second one maximizes the number of generated “artificial” CS that are not covered by solutions (detection rules) of the first population [S41].

\subsubsection{Manual approaches}
\label{subsubsec:Manualapproaches}
Manual techniques are human-centric, tedious, time-consuming, and error prone. These techniques require a great human effort, therefore not effective for detecting CS in large systems. According to the authors of \cite{Kessentini2014}, another important issue is that locating CS manually has been described as more a human intuition than an exact science.

The only study that uses a manual approach is [S52], where a catalog for the detection and handling of model smells for MATLAB / Simulink is presented. In this study, 3 types of techniques are used - manual, metric-based, symptom-based - according to the type of smell. The authors note that the detection of certain smells like the \textit{Vague Name} or \textit{Non-optimal Signal Grouping} can only be performed by manual inspection, because of the expressiveness of the natural language.

\subsection{Dataset availability (\textbf{F2})}
\label{subsec:datasetavailability}

The second finding is whether the underlying dataset is available - a precondition for study replication. When we talk about the dataset, i.e. the oracle, we are considering the software systems where CS and anti-patterns were detected, the type and number of CS and anti-patterns detected, and other data needed for the method used, e.g. if it is a metric-based approach the dataset must have the metrics for each application.

Only 12 studies present the available dataset, providing a link to it, however 2 studies, [S28] and [S32], no longer have the active links. Thus, only 12.0\% of the studies (10 out of 83, [S18, S27, S38, S51, S56, S59, S69, S70, S74, S82]) provide the dataset.

Another important feature for defining the dataset is which software systems are used in studies on which CS detection is performed. The number of software systems used in each study varies widely, and there are studies ranging from only one system to studies using 74 Java software systems and 184 Android apps with source code hosted in open source repositories. Most studies (83.1\%) use open-source software. Proprietary software is used in 3.6\% of studies and the use of the two types, open-source and proprietary, is used in 3.6\% of studies. It should be noted that 9.7\% of studies do not make any reference to the software systems being analysed.

\begin{table}[htpb]
\caption{Top ten open-source software projects used in the studies}
\label{table:open-sourcesoftwareprojects} \centering
\smallskip
\begin{tabular}{|l|c|c|}
 \hline
 \textbf{Open-source software} & \textbf{Nº of Studies} & \textbf{\% Studies}\\
 \hline
 Apache Xerces & 28 & 33.7\% \\ \hline
 GanttProject & 14 & 16.9\% \\ \hline
 ArgoUML & 11 & 13.3\% \\ \hline
 Apache Ant & 10 & 12.0\% \\ \hline
 JFreeChart & 8 & 9.6\% \\ \hline
 Log4J & 7 & 8.4\% \\ \hline
 Azureus & 7 & 8.4\% \\ \hline
 Eclipse & 7 & 8.4\% \\ \hline
 JUnit & 5 & 6.0\% \\ \hline
 JHotDraw & 5 & 6.0\% \\ \hline
\end{tabular}
\end{table}

Table \ref{table:open-sourcesoftwareprojects} presents the most used open-source software in the studies, as well as the number of studies where they are used and the overall percentage. Apache Xerces is the most used (33.7\% of the studies), followed by GanttProject with 16.9\%, ArgoUML with 13.3\%, and Apache Ant used in 12.0\% of the studies.

\subsection{Programming language (\textbf{F3})}
\label{subsec:programminglanguage}
\begin{figure}[!ht]
 \centering
 {\epsfig{file = 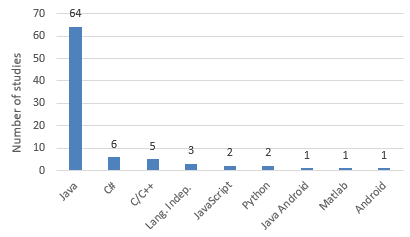, width = 8.4cm}}
 \caption{Programming languages and number of studies that use them}
 \label{fig:Program-Languages}
\end{figure}

In our research we do not make any restriction regarding the object-oriented programming language that supports the detection of CS. So, we have CS detection in 7 types of languages, in addition to the techniques that are language independent (3 studies) and a study [S66] that is for Android Apps without defining the type of language, as shown in Figure \ref{fig:Program-Languages} 

77.1\% of the studies (64 out of 83) use Java as a target language for the detection of CS, and therefore this is clearly the most used one.  C\# is the second most used programming language, with 6 studies (7.2\%), the third most used language is C/C++ with 5 studies (6.0\%). JavaScript and Python are used in 2 studies (2.4\%). Finally, we have 2 languages, MatLab and Java Android, which are used in only 1 study (1.2\%). In total we found seven different types of program languages to be used as support for the detection of CS.

In our analysis we found that 3.6\% of studies (3 out of 83, [S20, S32, S47]) present language-independent CS detection technique.
When we related the studies that are language-independent with the used approach, we found that two of the three studies used Symptom-based and one the Metric-based approach. These results are in line with what was expected, since a symptom-based approach is the most susceptible of being adapted to different programming languages.

\begin{figure}[!ht]
 \centering
 {\epsfig{file = 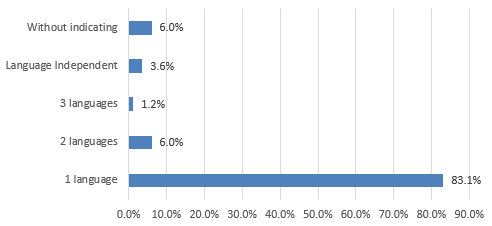, width = 8.4cm}}
 \caption{Number of languages used in each study}
 \label{fig:Num-Languages-study}
\end{figure}

Multi-language support for CS detection is also very limited as shown in Figure \ref{fig:Num-Languages-study}.
In addition to the three independent language studies, only [S28], one study (1.2\%) of the 83 analyzed, supports 3 programming languages. Five studies (6.0\%, [S3, S9, S12, S15, S57]) detect CS in 2 languages and 69 studies (83.1\%) only detect in a one programming language. Five studies (6.0\%) explain the detection technique, but do not refer to any language.

When we analyze the 5 studies that do not indicate any programming language, we find that all use a visualization-based approach, i.e., 41.7\% (5 out of 12) of studies that use visualization techniques do not indicate any programming language.

\subsection{Code smells detected (\textbf{F4})}
\label{subsec:Codesmellsdetection}
Several authors use different names for the same CS, so to simplify the analysis we have grouped the different CS with the same mean into one, for example, \textit{Blob}, \textit{Large Class} and \textit{God Class} were all grouped in \textit{God Class}. The description of CS can be found in Appendix \ref{AppendixD}.

In Table \ref{table:Code-smells-detected} we can see the CS that are used in more than 3 studies, the number of studies in which they are detected and the respective percentage. 
As we have already mentioned in subsection \ref{subsec:programminglanguage}, in this systematic review we do not make any restriction regarding the Object Oriented programming language used. Thus, considering all Object Oriented programming languages 68 different CS are detected, much more than the 22 described by Fowler \cite{Fowler1999}. \textit{God Class} is the most detected CS, being used in 51.8 \% of the studies, followed by \textit{Feature Envy} and \textit{Long Method} with 33.7 \% and 26.5 \%, respectively.

\begin{table}[htpb]
\caption{Code smells detected in more than 3 studies}
\label{table:Code-smells-detected} \centering
\smallskip
\begin{tabular}{|p{4.0cm} |c|c|}
 \hline 
 \textbf {Code smell} & \textbf{Nº of studies} & \textbf{\% Studies}\\
 \hline
God Class (Large Class or Blob) & 43 & 51.8\% \\ \hline
Feature Envy & 28 & 33.7\% \\ \hline
Long Method & 22 & 26.5\% \\ \hline
Data class & 18 & 21.7\% \\ \hline
Functional Decomposition & 17 & 20.5\% \\ \hline
Spaghetti Code & 17 & 20.5\% \\ \hline
Long Parameter List & 12 & 14.5\% \\ \hline
Swiss Army Knife & 11 & 13.3\% \\ \hline
Refused Bequest & 10 & 12.0\% \\ \hline
Shotgun Surgery & 10 & 12.0\% \\ \hline
Code clone/Duplicated code & 9 & 10.8\% \\ \hline
Lazy Class & 8 & 9.6\% \\ \hline
Divergent Change & 7 & 8.4\% \\ \hline
Dead Code & 4 & 4.8\% \\ \hline
Switch Statement & 4 & 4.8\% \\ \hline
\end{tabular}
\end{table}

All 68 CS detected are listed in Appendix \ref{AppendixE}, as well as the number of studies in which they are detected, their percentage, and the programming languages in which they are detected.

When we analyzed the CS detected in each study, we found that the number is low, with an average of 3.6 CS per study and a mode of 1 CS. Only the study [S57], with a symptom-based approach, detects the 22 CS described by Fowler \cite{Fowler1999}.

\begin{figure}[!ht]
 \centering
 {\epsfig{file = 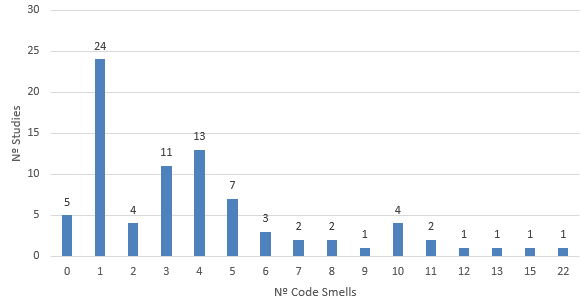, width = 8cm}}
 \caption{Number of code smells detected by number of studies }
 \label{fig:CS-per-study}
\end{figure}

Figure \ref{fig:CS-per-study} shows the number of CS detected by number of studies. We can see that the number of smells most detected is 1 (in 24 studies), 4 smells are detected in 13 studies and 11 studies detect 3 smells. The detection of 12, 13, 15 and 22 smells is performed in only 1 study.
It should be noted that 5 studies do not indicate which CS they detected. It is also important to note that these 5 studies use a visualization-based approach.

\subsection{Machine Learning techniques used (\textbf{F5})}
\label{subsec:MachineLearningtechniques }
ML algorithms present many variants and different parameter configurations, making it difficult to compare them. For example, the Support Vector Machines algorithm has variants such as SMO, LibSVM, etc. Decision Trees can use various algorithms such as C5.0, C4.5 (J48), CART, ID3, etc. As algorithms are presented with different details in the studies, for a better understanding of the algorithm used, we classify ML algorithms in their main category, creating 9 groups as shown in the table \ref{table:ML-algorithms-used}. 

Table \ref{table:ML-algorithms-used} shows the ML algorithms, the number of studies using the algorithm and its percentage, as well as the ID of the studies that use the algorithm.

\begin{table}[htpb]
\caption{ML algorithms used in the studies}
\label{table:ML-algorithms-used} \centering
\smallskip
\begin{tabular}{|p{2.5cm}|p{1cm}|p{1.1cm}|p{2.3cm}|}
 \hline
 \textbf{ML algorithm} & \textbf{Nº of Studies} & \textbf{\% Studies} & \textbf{Studies}\\
 \hline
Genetic Programming             & 9 & 10.8\% & S31, S41, S43, S45, S58, S64, S66, S70, S79 \\ \hline
Decision Tree                   & 8 & 9.6\%  & S6, S28, S36, S40, S53, S56, S77, S83       \\ \hline
Support Vector Machines (SVM)   & 6 & 7.2\%  & S33, S34, S36, S56, S71, S77                \\ \hline
Association Rules               & 6 & 7.2\%  & S36, S42, S50, S54, S56, S77                \\ \hline
Bayesian Algorithms             & 5 & 6.0\%  & S18, S27, S36, S56, S77                     \\ \hline
Random Forest                   & 3 & 3.6\%  & S36, S56, S77                               \\ \hline
Neural Network                  & 2 & 2.4\%  & S74, S75                                    \\ \hline
Regression models               & 1 & 1.2\%  & S22                                         \\ \hline
Artificial Immune Systems (AIS) & 1 & 1.2\%  & S24                                         \\ \hline
\end{tabular}
\end{table}

From the 83 primary studies analyzed 35\% of the studies (29 out of 83, [S6, S18, S22, S24, S27, S28, S31, S33, S34, S36, S40, S41, S42, S43, S45, S50, S53, S54, S56, S58, S64, S66, S70, S71, S74, S75, S77, S79, S83]) use ML techniques in CS detection. Except for 3 studies [S36, S56, S77] where multiple ML algorithms are used, all 26 other studies use only 1 algorithm in the CS detection.

The most widely used algorithms are Genetic algorithms (9 out 83, [S31, S41, S43, S45, S58, S64, S66, S70, S79]) and Decision Trees (8 out 83, [S6, S28, S36, S40, S53, S56, S77, S83]), which are used in 10.8\% and 9.6\%, respectively, of the analyzed studies.
We think that a possible reason why genetic algorithms are the most used algorithm, is because they are used to generate the CS detection rules and to find the best threshold values to be used in the detection rules.
Regarding Decision trees, it is due to the easy interpretation of the models, mainly in its variant C4.5 / J48 / C5.0.

The third most used algorithms for ML, used in 7.2\% of the studies, are Support Vector Machines (SVM) (6 out 83, [S33, S34, S36, S56, S71, S77]) and association rules (6 out 83, [S36, S42, S50, S54, S56, S77]) with Apriori and JRip being the most used.

Bayesian Algorithms are the fifth most used algorithm with 6.0\% (5 out 83, [S18, S27, S36, S56, S77]). 

The other 4 ML algorithms that were also used are Random Forest (in 3 studies), Neural Network (in 2 studies), Regression models (in 1 study) and Artificial Immune Systems (AIS) (in 1 study).

\subsection{Evaluation of techniques (\textbf{F6})}
\label{subsec:Evaluationtechniques}
The evaluation of the technique used is an important factor to realize its effectiveness and consequently choose the best technique or tool.
The main metrics used to evaluate the techniques are accuracy, precision, recall, F-measure. These 4 metrics are calculated based on true positives (TP), false positives (FP), false negatives (FN), and true negatives (TN) instances of CS detected, according to the following formulas:
\begin{itemize}
 \item Accuracy = (TP + TN) / (TP + FP + FN + TN)
 \item Precision = TP / (TP + FP)
 \item Recall = TP / (TP + FN)
 \item F-measure = 2 * (Recall * Precision) / (Recall + Precision)
\end{itemize}
In the 83 articles analyzed, 86.7\% (72 studies) evaluated the technique used and 13.3\% (11 studies) did not evaluate the technique. 
Table \ref{table:metrics-evaluate-techniques} shows the most used evaluation metrics.Precision is the most used with 46 studies (55.4\%), followed by recall in 44 studies (53.0\%) and F-measure in 17 studies (20.5\%). It should be noted that 28 studies (33.7 \%) use other metrics for evaluation such as the number of detected defects, Area Under ROC, Standard Error (SE) and Mean Square Error (MSE), Root Mean Squared Prediction Error (RMSPE), Prediction Error (PE), etc.

\begin{table}[htpb]
\caption{Metrics used to evaluate the detection techniques}
\label{table:metrics-evaluate-techniques} \centering
\smallskip
\begin{tabular}{|l|c|c|}
 \hline 
 \textbf {Metric} & \textbf{Nº of studies} & \textbf{\% Studies}\\
 \hline
Precision & 46 & 55.4\% \\ \hline
Recall & 44 & 53.0\% \\ \hline
F-measure & 17 & 20.5\% \\ \hline
Accuracy & 10 & 12.0\% \\ \hline
Other & 28 & 33.7\% \\ \hline
Without evaluation & 11 & 13.3\% \\ \hline
\end{tabular}
\end{table}

In the last years the most used metrics in the evaluation are the precision and recall, but until 2010 few studies have evaluations based on these metrics, presenting only the CS detected.
When we analyze the evaluations of the different techniques, we verified that the results depend on the applications used to create the oracle and the CS detected, so we have several studies that have chosen to present the means of precision and recall.

Regarding the different approaches used, we can conclude that:
\begin{enumerate}[noitemsep]
\item in manual approaches and cooperative-based approaches, since we only have one study for each, we cannot draw conclusions;
\item in the visualization-based approaches, most of the evaluations presented are qualitative, and almost half of the studies do not present an evaluation;
\item in relation to the other 4 approaches (probabilistic, metric, symptom-based, and search-based), all present at least one study/technique with 100\% recall and precision results.
\item It is difficult to make comparisons between the different techniques, since, except for the studies of the same author(s), all the others have different oracles.
\end{enumerate}

The usual way to build an oracle is to choose a set of software systems (typically open source), choose the CS that you want to detect and ask a group of MSc students (3, 4 or 5 students), supervised by experts (e.g. software engineers), to identify the occurrences of smells in systems. In case of doubt of a candidate to CS, either or the expert decides, or the group reaches to a consensus on whether this candidate is or not a smell. As you can see, the creation of an oracle is not an easy task, because it requires a tremendous amount of manual work in the detection of CS, having all the problems of a manual approach (see section \ref{subsubsec:Manualapproaches}) mainly the subjectivity.

For a rigorous comparison of the evaluation of the different techniques, it is necessary to use common datasets and oracles (see section \ref{subsec:datasetavailability}), which does not happen today.

\subsection{Detection tools (\textbf{F7})}
\label{subsec:Detectiontools}
Comparing the results of CS detection tools is important to understand the performance of the techniques associated with the tool and consequently to know which one is the best. It is also important to create tools that allow us to replicate studies.

When we analyzed which studies created a detection tool, we found that 61.4\% (51 out of 83) studies developed a tool, as show in table \ref{table:developedtoolapproach}.

\begin{table}[htpb]
\caption{Number of studies that developed a tool and its approach}
\label{table:developedtoolapproach} \centering
\smallskip
\begin{tabular}{|m{2cm}|m{0.9cm}|m{1cm}|m{1.3cm}|m{1.1cm}|}
 \hline
 \textbf{Approaches} & \textbf{Nº studies} & \textbf{Nº studies with tool} & \textbf{\% Studies in the approach} & \textbf{\% Studies in total}\\
 \hline
 Symptom-based & 16 & 13 & 81.3\% & 15.7\%\\ \hline
 Metric-Based & 20 & 13 & 65.0\% & 15.7\%\\ \hline
 Visualization-based & 12 & 10 & 83.3\% & 12.0\%\\ \hline
 Search-Based & 25 & 9 & 36.0\% & 10.8\%\\ \hline
 Probabilistic & 10 & 6 & 60.0\% & 7.2\%\\ \hline
 Cooperative-based & 1 & 0 & 0.0\% & 0.0\%\\ \hline
 Manual & 1 & 0 & 0.0\% & 0.0\%\\ \hline
\end{tabular}
\end{table}

The symptom-based and metrics-based approaches are those that present the most tools developed with 15.7\% (13 out of 83 studies), follow by Visualization-based  with 12.0\% (10 out of 83 studies) (see Table \ref{table:developedtoolapproach}).
On the opposite side, there is the Probabilistic approach where only 7.2\% (6 out of 83 studies) present developed tools. 

When we analyze the percentage of studies that develop tools within each approach, we find that visualization-based and symptom-based approaches are those that have a greater number of developed tools with 83.3\% (10 out of 12 studies) and 81.3\% (13 out of 16 studies), respectively (see Table \ref{table:developedtoolapproach}).

On the other side, there is the search-based approach where only 36.0\% (9 out of 25 studies) present developed tools.
In this approach, less than half of the studies present a tool because they chose to use already developed external tools instead of creating new ones. For example, some studies [S34, S36, S56, S77] use Weka \footnote{Weka is a collection of ML algorithms for data mining tasks (www.cs.waikato.ac.nz/ml/weka/)} to implement their techniques.

As Rasool and Arshad mentioned in their study \cite{Rasool2015}, it becomes arduous to find common tools that performed experiments on common systems for extracting common smells. Different techniques perform experiments on different systems and present their results in different formats.
When analyzing the results of different tools to verify their results, examining the same software packages and CS, we verified a disparity of results \cite{Fernandes2016,Rasool2015}.

\subsection{Thresholds definition (\textbf{F8})}
\label{subsec:Thresholdsdefinition}
Threshold values are a very important component in some detection techniques because they are the values that define whether or not a candidate is a CS. Its definition is very complicated and one of the reasons there is so much disparity in the detection results of CS (see section \ref{subsec:Detectiontools}). Some studies use genetic algorithms to calibrate threshold values as a way of reducing subjectivity in CS detection, e.g. [S70].

\begin{table}[htpb]
\caption{Number of studies that use thresholds in CS detection}
\label{table:usethresholds} \centering
 \begin{tabular}{|m{2cm}|m{0.9cm}|m{1cm}|m{1.3cm}|m{1.1cm}|}
 \hline
  \textbf{Approaches} & \textbf{Nº studies} & \textbf{Nº use thresholds} & \textbf{\% Studies in the approach} & \textbf{\% Studies in total}\\
 \hline
 Metric-Based & 20 & 15 & 75.0\% & 18.1\%\\ \hline
 Symptom-based & 16 & 11 & 68.8\% & 13.3\%\\ \hline
 Search-Based & 25 & 8 & 32.0\% & 9.6\%\\ \hline
 Probabilistic & 10 & 8 & 80.0\% & 9.6\%\\ \hline 
 Visualization-based & 12 & 1 & 8.3\% & 1.2\%\\ \hline 
 Cooperative-based & 1 & 1 & 100.0\% & 1.2\%\\ \hline
 Manual & 1 & 0 & 0.0\% & 0.0\%\\ \hline
\end{tabular}
\end{table}

A total of 44 papers use thresholds in their detection technique, representing 53.0\% of all studies. 47.0\% of studies (39 out of 83 studies) did not use thresholds.

Without the Cooperative-based approach (which presents only 1 study), in the total of studies, metric-based and symptom-based approaches are those that present the most tools developed with 18.1\% (15 out of 83 studies) and 13.3\% (11 out of 83 studies), respectively (see Table \ref{table:usethresholds}).

When we analyze the number of studies within each approach that uses thresholds,we find that the three approaches that most use thresholds in their detection techniques are Probabilistic with 80\% (8 out of 10 studies), Metric-based with 75.0\% (15 out of 20 studies), and Symptom-based with 68.8\% (11 out of 16 studies). In visualization-based approaches, only one study use thresholds in their CS detection techniques, as shown in Table \ref{table:usethresholds}.

Analyzing the detection techniques, we verified that these results are in line with what was expected, since we found that the probabilistic and metric-based approaches are those that most need to use thresholds. In the probabilistic approaches to define the values of support, confidence and probabilistic decision values. In metric-based approaches, it is essential to define threshold values for the different metrics that compose the rules.

\subsection{Validation of techniques (\textbf{F9})}
\label{subsec:Validationtechniquess}
The validation of a technique is performed by comparing the results obtained by the technique, with the results obtained through another technique with similar objectives. Obviously, both techniques must detect the same CS in the same software systems.
The most usual forms of validation are: using the techniques of various existing approaches, such as manuals; use existing tools; comparing the results with those of other published papers.

When we analyze how many studies are validating their technique (see Table \ref{table:Toolsvalidation}), we verified that 62.7\% (52 out of 83, [S3, S6, S8, S13, S15, S18, S20, S23, S24, S26, S27, S28, S30, S31, S33, S34, S38, S41, S42, S43, S44, S45, S47, S48, S49, S50, S51, S53, S54, S55, S56, S58, S59, S60, S61, S62, S64, S65, S66, S67, S68, S70, S71, S72, S73, S74, S75, S77, S78, S79, S81, S83]) perform validation. In opposition 37.3\% (31 out of 83, [S1, S2, S4, S5, S7, S9, S10, S11, S12, S14, S16, S17, S19, S21, S22, S25, S29, S32, S35, S36, S37, S39, S40, S46, S52, S57, S63, S69, S76, S80, S82]) of the studies do not validate the technique.

\begin{table}[htpb]
\caption{Tools / approach used by the studies for validation}
\label{table:Toolsvalidation} \centering
\smallskip
\begin{tabular}{|m{2cm}|m{1.2cm}|c|m{2.3cm}|}
 \hline
 \textbf{Tool/approach} & \textbf{Nº of Studies} & \textbf{\% Studies} & \textbf{Studies}\\
 \hline
  DECOR & 14 & 26.9\% & S18, S24, S27, S30, S31, S42, S43, S45, S48, S50, S51, S59, S62, S79 \\ \hline
Manually & 14 & 26.9\% & S3, S6, S8, S13, S15, S23, S31, S38, S43, S44, S53, S54, S56, S66 \\ \hline
JDeodorant & 10 & 19.2\% & S42, S44, S45, S50, S51, S54, S62, S72, S73, S75 \\ \hline
 iPlasma & 8 & 15.4\% & S26, S49, S53, S54, S56, S65, S68, S81 \\ \hline
 Machine Learning & 7 & 13.5\% & S24, S43, S58, S66, S67, S79, S83 \\ \hline
 Papers & 6 & 11.5\% & S41, S51, S60, S61, S74, S77 \\ \hline
 DETEX & 3 & 5.8\% & S33, S34, S71 \\ \hline
 Incode & 3 & 5.8\% & S53, S64, S44 \\ \hline
 inFusion & 3 & 5.8\% & S65, S53, S49 \\ \hline
 PMD & 3 & 5.8\% & S49, S56, S53 \\ \hline
 BDTEX & 2 & 3.8\% & S33, S59 \\ \hline
 CodePro AnalytiX & 2 & 3.8\% & S55, S78 \\ \hline
 Jtombstone & 2 & 3.8\% & S55, S78 \\ \hline
 Rule Marinescu & 2 & 3.8\% & S47, S65 \\ \hline
 AntiPattern Scanner & 1 & 1.9\% & S56 \\ \hline
 Bellon benchmark & 1 & 1.9\% & S15 \\ \hline
 Checkstyle & 1 & 1.9\% & S49 \\ \hline
 DCPP & 1 & 1.9\% & S50 \\ \hline
 DUM-Tool & 1 & 1.9\% & S78 \\ \hline
 Essere & 1 & 1.9\% & S68 \\ \hline
 Fluid Tool & 1 & 1.9\% & S56 \\ \hline
 HIST & 1 & 1.9\% & S42 \\ \hline
 JADET & 1 & 1.9\% & S20 \\ \hline
 Jmove & 1 & 1.9\% & S75 \\ \hline
 JSNOSE & 1 & 1.9\% & S70 \\ \hline
 Ndepend & 1 & 1.9\% & S73 \\ \hline
 NiCad & 1 & 1.9\% & S28 \\ \hline
 SonarQube & 1 & 1.9\% & S50 \\ \hline
\end{tabular}
\end{table}

Considering the differences between techniques and all subjectivity in a technique (see sections \ref{subsec:Thresholdsdefinition}, \ref{subsec:Detectiontools}, \ref{subsec:Evaluationtechniques}), we can conclude that it is not easy to perform validations with tools that implement other techniques, even if they have the same goals. Thus, it is not surprising that one of the most common method of validating the results is manually, with a percentage of 26.9\% of the studies (14 of the 52 studies doing validation, [S3, S6, S8, S13, S15, S23, S31, S38, S43, S44, S53, S54, S56, S66]), as shown in Table \ref{table:Toolsvalidation}.

Some authors as in [S8] claim that, validation was performed manually because only maintainers can assess the presence of defects in design depending on their design choices and in the context, or as in [S23] where validation is performed by independent engineers who assess whether suspicious classes are smells, depending on the contexts of the systems.

Equally with manual validation is the use of the DECOR tool \cite{Moha2010a}, also used in 26.9\% of studies (14 out of 52, [S18, S24, S27, S30, S31, S42, S43, S45, S48, S50, S51, S59, S62, S79]), this approach is based on symptoms. DECOR is a tool proposed by Moha et al. \cite{Moha2010a} which uses domain-specific language to describe CS. They used this language to describe well-known smells, \textit{Blob} (aka \textit{Long Class}), \textit{Functional Decomposition}, \textit{Spaghetti Code}, and \textit{Swiss Army Knife}. They also presented algorithms to parse rules and automatically generate detection algorithms.

The following two tools most used in validation, with 19.2\% (10 out of 52 studies) are JDeodorant \cite{Fokaefs2007} used for validation of the studies [S42, S44, S45, S50, S51, S54, S62, S72, S73, S75], and iPlasma \cite{Marinescu2005} for the studies [S26, S49, S53, S54, S56, S65, S68, S81]. JDeodorant \footnote{https://users.encs.concordia.ca/~nikolaos/jdeodorant/} is a plug-in for eclipse developed by Fokaefs et al. for automatic detection of CS (\textit{God Class}, \textit{Type Check}, \textit{Feature Envy}, \textit{Long Method}) and performs refactoring. iPlasma \footnote{http://loose.utt.ro/iplasma/} is a tool that uses a metric-based approach to CS detection developed by Marinescu et al.

Seven studies [S24, S43, S58, S66, S67, S79, S83] compare their results with the results obtained through ML techniques, namely Genetic Programming (GP), BBN, and Support Vector Machines (SVM). The ML techniques represent 13.5\% of the studies (7 out of 52) that perform validation.

As we can see in the table \ref{table:Toolsvalidation}, where we present the 28 different ways of doing validation, there are still many other tools that are used to validate detection techniques.

\subsection{Replication of the studies (\textbf{F10})}
\label{subsec:Replicationstudies}
The replication of a study is an important process in software engineering, and its importance is highlighted by several authors such as Shull et al. \cite{Shull2008} and Barbara Kitchenham \cite{Kitchenham2008}.
According to Shull et al. \cite{Shull2008}, the replication helps to ``\textit{better understand software engineering phenomena and how to improve the practice of software development. One important benefit of replications is that they help mature software engineering knowledge by addressing both internal and external validity problems.}". The same authors also mention that in terms of external validation, replications help to generalize the results, demonstrating that they do not depend on the specific conditions of the original study. In terms of internal validity, replications also help researchers show the range of conditions under which experimental results hold. These authors still identify two types of replication: exact replications and conceptual replications.

Another author to emphasize the importance of replication is Kitchenham \cite{Kitchenham2008}, claiming that "replication is a basic component of the scientific method, so it hardly needs to be justified."

Given the importance of replication, it is important that the studies provide the necessary information to enable replication. Especially in exact replications, where the procedures of an experiment are followed as closely as possible to determine if the same results can be obtained \cite{Shull2008}. Thus, our goal is not to perform replications, but to verify that the study has the conditions to be replicated.

According to Carver \cite{carver2010towards} \cite{Carver2014}, a replication paper should provide the following information about the original study (at a minimum): Research questions, Participants, Design, Artifacts, Context variables, Summary of results.

This information about the original study is necessary to provide sufficient context to understand replication. Thus, we consider that for a study to be replicated, it must have available this information identified by Carver.

Oracles are extremely important for the replication of CS detection studies. The building of oracles is one of the methods that more subjectivity causes in some techniques of detection, since they are essentially manual processes, with all the inherent problems (already mentioned in previous topics).
As we have seen in section \ref{subsec:datasetavailability}, only 10 studies present the available dataset, providing a link to it, however 2 studies, [S28] and [S32], no longer have the active links. Thus, only 12.0\% of the studies (10 out of 83, [S18, S27, S38, S51, S56, S59, S69, S70, S74, S82]) provide the dataset, and are candidates for replication.

Another of the important information for the replication is the existence of an artifact, it happens that the studies [S70] amd [S74] does not present an artifact, therefore it cannot be replicated.

We conclude that only 9.6\% of the studies (8 out of 83, [S18, S27, S38, S51, S56, S59, S69, S82]) can be replicated because they provide the information claimed by Carver \cite{carver2010towards}. 

It is noteworthy that [S51] makes available on the Internet a replication package composed of Oracles, Change History of the Object systems, Identified Smells, Object systems, Additional Analysis - Evaluating HIST on Cassandra Releases.

\subsection{Visualization techniques (\textbf{F11})}
\label{subsec:Visualizationtechniques}
The CS visualization can be approached in two different ways, (i) the CS detection is done through a non-visual approach and the visualization being performed to show the CS in the code, (ii) the CS detection is performed through a visual approach.

Regarding the first approach, the visualization is only to show previously detected CS, by a non-visualization approach, we found 5 studies [S9, S40, S57, S72, S81], corresponding to 6.0\% of the studies analyzed in this SLR. Thus, we can conclude that most studies are only dedicated to detecting CS, but do not pay much attention to visualization. Most of the proposed CS visualization shows the CS inside the code itself. This approach works for some systems, but when we are in the presence of large legacy systems, it is too detailed for a global refactoring strategy. Thus, a more macro approach is required, without losing detail, to present CS in a more aggregated form. 

In relation to the second approach, where a visualization-based approach is used to detect CS, it represents 14.5\% of the studies (12 out of 83, [S1, S2, S7, S11, S12, S16, S17, S19, S21, S39, S46, S76]). One of the problems pointed to the visualization-based approach is the scalability for large systems, since this type of approach is semi-automatic, requiring human intervention. In relation to this aspect, we found only 3 studies [S7, S17, S16] with solutions dedicated to large systems.

Most studies do a visualization showing the system structure in packages, classes, and methods, but it is not enough, it is necessary to adapt the views according to the type of CS, and this is still not generalized. The focus must be the CS and not the software structure, for example, it is not necessary to show the parts of the software where there are no smells, since it is only adding data to the views, complicating them without adding information.

Combining the two types of approaches, we conclude that 20.5\% of the studies (17 out of 83) use some kind of visualization in their approach.

\begin{figure*}[!ht]
 \centering
 {\epsfig{file = 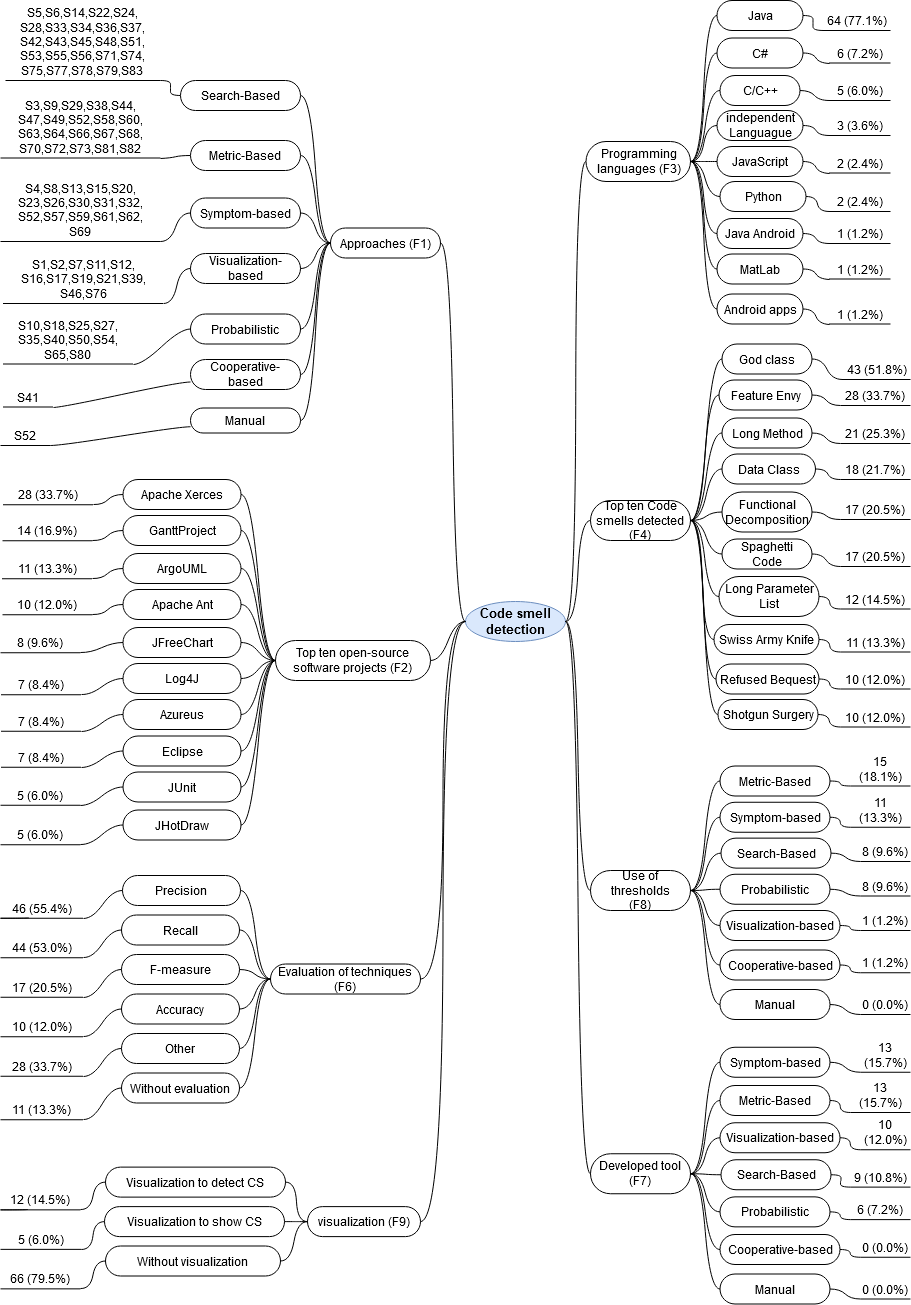, width = 15.5cm}}
  \caption{Summary of main findings}
 \label{fig:Findings-resume}
\end{figure*}

As for code smells coverage, the \textit{Duplicated Code} CS (aka \textit{Code Clones}) is definitely the one more where visualization techniques have been applied more intensively. Recall from Section \ref{sec:Relatedwork} that Zhang et al. \cite{Zhang2010} systematic review on CS revealed that \textit{Duplicated Code} is the most widely studied CS. Also in that section, we referred to Fernandes et al. \cite{Fernandes2016} systematic review that concluded that \textit{Duplicated Code} was among the top-three CS detected by tools, due to its importance.
The application of visualization to the \textit{Duplicated Code} CS ranges from 2D techniques (e.g. dot plots / scatterplots, wheel views / chord diagrams, and other graph-based and polymetric view-based ones) to more sophisticated techniques, such as those based on 3D metaphors and virtual reality.  A comprehensive mapping study on the topic of \textit{Duplicated Code} visualization has just been published \cite{Hammad2020}.

\section{Discussion}
\label{sec:discussion}
We now address our research questions, starting by discussing what we found for each of the research questions, mainly the benefits and limitations of evidence of these findings. The mind map in Figure \ref{fig:Findings-resume} provides a summary of main findings. Finally, we discuss the validation and limitations of this systematic review.

\subsection{Research Questions (\textbf{RQ})}
\label{subsec:Researchquestions}
\begin{figure}[!ht]
 \centering
 {\epsfig{file = 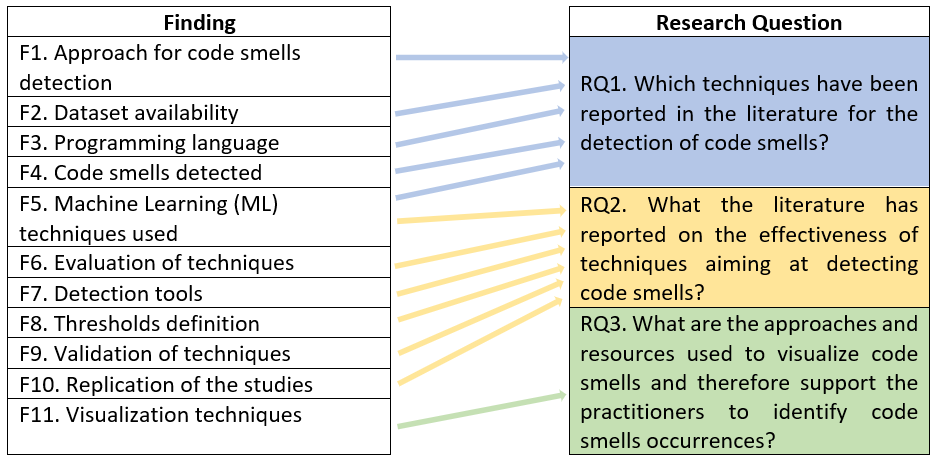, width = 8.4cm}}
 \caption{Relations between findings and research questions}
 \label{fig:Findings-rq}
\end{figure}

This subsection aims to discuss the answers to the three research questions and how the findings and selected documents addressed these issues. In figure \ref{fig:slrfinal} we show the selected studies and the respective research questions they focus on. Regarding how findings (\textbf{F}) interrelate with research questions, findings F1, F2, F3, F4, F5 support the answer of RQ1, findings F5, F6, F7, F8, F9, F10, support the answer of RQ2, and, finally, F11 supports the answer of RQ3 (see  figure \ref{fig:Findings-rq}).

\vspace{5pt}
\textbf{RQ1}. \textit{Which techniques have been reported in the literature for  the detection of CS?} 

To answer this research question, we classified the detection techniques in seven categories, following the classification presented in Kessentini et al. \cite{Kessentini2014} (see F1, subsection \ref{subsec:ApproachCSdetection}). 
The search-based approach is applied in 30.1\% studies. These types of approaches are inspired by contributions in the domain of Search-Based Software Engineering (SBSE) and most techniques in this category apply ML algorithms, with special incidence in the algorithms of genetic programming, decision tree, and association rules. The second most used approach is the metric-based applied in 24.1\% studies. The metric-based approach consists in the use of rules based on a set of metrics and respective thresholds to detect specific CS. The third most used approach, with 19.3\% of studies, is symptom-based approach. It consists of techniques that describe the symptoms of CS and later translate these symptoms into detection algorithms.

Regarding datasets (see F2, subsection \ref{subsec:datasetavailability}), more specifically the oracles used by the different techniques, we conclude that only 10 studies (12.0\% of studies) provide the same. Most studies, 83.1\%, use open-source software, the most used systems being Apache Xerces (33.7\% of studies), GanttProject (16.9\%), followed by ArgoUML with 13.3\%.

Java is used in 64 studies, i.e. 77.1\% of the studies use Java as a support language for the detection of CS. C\# and C++ are the other two most commonly used programming languages that support CS detection, being used in 7.2\% and 6.0\% of studies respectively (see F3, subsection \ref{subsec:programminglanguage}). Multi-language support for code detection is also very limited, with the majority (83.1\%) of the studies analyzed supporting only one language. The maximum number of supported languages is three, being reported exclusively in [S28].

Regarding the most detected CS, \textit{God Class} stands out with 51.8\% (see F4, subsection \ref{subsec:Codesmellsdetection}). \textit{Feature Envy} and \textit{Long Method} with 33.7\% and 26.5\%, respectively, are the two most commonly used CS. When we analyze the number of CS detected by each study, we found that they detected on average three CS, but the most frequent is the studies detect only 1 CS.

\vspace{5pt}
\textbf{RQ2}. \textit{What literature has reported on the effectiveness of techniques aiming at detecting CS?} 

Finding out the most effective technique to detect CS is not a trivial task. We have realized that the identified approaches have all pros and cons, presenting factors that can bias the results.
Although, in the evaluation of the techniques (F6, subsection \ref{subsec:Evaluationtechniques}) we verified that there are 4 approaches (probabilistic, metric-based, symptom-based, and search-based) that present techniques with 100\% accuracy and recall results, the detection problem is very far to be solved.
These results only apply to the detection of simpler CS, e.g. \textit{God Class}, so it is not surprising that 51.8\% of the studies use this CS, as we can see in table \ref{table:Code-smells-detected}. In relation to the more complex CS, the results are much lower and very few studies use them. We cannot forget that only one study [S57] detects the 22 CS described by Fowler \cite{Fowler1999}.

The answer to RQ2 is that, there is no one technique, but several techniques, depending on several factors such as:
\begin{itemize}
 \item Code smell to detect - We found that there is no technique that generalizes to all CS. When we analyze the studies that detect the greatest number of CS, [S38, S53, S57, S63, S69] are the studies that detect more than 10 CS and make an evaluation, we find that precision and recall depend on smell and there are large differences.
 \item Software systems - The same technique when detecting the same CS in different software systems, there is a great discrepancy in the number of false positives and false negatives and consequently in precision and recall.
 \item Threshold values - There is no consensus regarding the definition of threshold values, the variation of this value causes more, or less, CS to be detected, thus varying the number of false positives. Some authors try to define thresholds automatically, namely using genetic programming algorithms.
 \item Oracle - There is no widespread practice of oracle sharing, few oracles are publicly available. Oracles are a key part of most CS detection processes, for example, in the training of ML algorithms.
\end{itemize}

Regarding the automation of CS detection processes, thus making them independent of thresholds, we found that 35\% of the studies used ML techniques. However, when we look at how many of these studies do not require thresholds, we find that only 18.1\% (15 out of 83, [S22, S33, S34, S36, S40, S43, S53, S56, S66, S71, S74, S75, S77, S79, S83]) are truly automatic.

\vspace{5pt}
\textbf{RQ3}. \textit{What are the approaches and resources used to visualize CS and therefore support the practitioners to identify CS occurrences?} 

The visualization and representation are of extreme importance, considering the variety of CS, possibilities of location in code (within methods, classes, between classes, etc.), and dimension of the code for a correct identification of the CS.
Unfortunately, most of the studies only detect it, does not visually represent the detected CS. 
In studies that do not use visualization-based approaches from 83 studies selected in this SLR, only five studies [S9, S40, S57, S72, S81] visually represent CS.

In the 14.5\% studies (12 out of 83) that use visualization-based approaches to detect CS, several methods are used to show the structure of the programs, such as: (1) city metaphors [S7, S17]; (2) 3D visualization technique [S16, S17]; (3) interactive ambient visualization [S19, S39, S46]; (4) multiple views adapted to CS [S12,S21]; (5) polymetric views [S2, S46]; (6) graph model [S1]; (7) Multivariate visualization techniques, such as parallel coordinates, and non-linear projection of multivariate data onto a 2D space [S76]; (8) in [S46] several views are displayed such as node-link-based dependency graphs, grids and spiral egocentric graphs, and relationship matrices.

With respect to large systems, only three studies present dedicated solutions.

\subsection{SLR validation}
\label{sec:SLRvalidation}
To ensure the reliability of the SLR, we carried out validations in 3 stages:

i) The first validation was carried out in the application of the inclusion and exclusion criteria, through the application of Fleiss' Kappa. Through this statistical measure, we validated the level of agreement between the researchers in the application of the inclusion and exclusion criteria.

ii) The second validation was carried out through a focus group, when the quality criteria were applied in stage 4.

iii) To validate the results of the SLR we conducted 3 surveys, with each survey divided into 2 parts, one on CS detection and another on CS visualization. Each question in the surveys consists of 3 parts: 1) the question about one of the findings that is evaluated on a 6 point Likert scale (Strong disagreement, Disagreement, Weak disagreement, Agreement, Strong agreement); 2) a slider between 0 and 4 that measures the degree of confidence of the answer; 3) an optional field to describe the justification of the answer or for comments.
The three inquiries were intended to: 1) Pre-test, with the aim of identifying unclear questions and collecting suggestions for improvement. The subjects chosen for the pre-test were Portuguese researchers with the most relevant work in the area of software engineering, totaling 27; 2) The subjects in the second survey were the authors of the studies that are part of this SLR, totaling 193; 3) The third survey was directed at the software visualization community; we chose the authors from all papers selected for the SLR on software visualization by Merino et al.\cite{MERINO2018} that were taken exclusively from the SOFTVIS and VISSOFT conferences, totaling 380; we also distributed this survey through a post on a Software Visualization blog\footnote{https://softvis.wordpress.com/}.  

The structure of the surveys, collected responses, and descriptive statistics on the latter are available at a github repository\footnote{https://github.com/dataset-cs-surveys/Dataset-CS-surveys.git}.

In table \ref{table:SummarySurvey} we present a summary of the results of the responses from this SLR' authors (2nd survey) and from the visualization community (3rd survey). As we can see, using the aforementioned scale, most participants agree with SLR results. The grayed cells in this table represent, for each finding, the answer(s) that obtained the highest score. We can then observe that: 10\% of the findings had \textit{Strong agreement} as its higher score, 80\% of the findings had \textit{Agreement} and, 20\% had \textit{Weak agreement}.

Regarding the question, \textit{Please select the 3 most often detected code smells?}, 
the answers placed the \textit{Long Method} as the most detected CS, followed by \textit{God Class} and \textit{Feature Envy}. In our SLR, based on actual data, we concluded that the most detected CS is \textit{God Class}, followed by \textit{Feature Envy} and \textit{Long Method}. This mismatch is small, since it only concerns the relative order of those 3 code smells, and shows that the community is well aware of which are the most often found ones.

\begin{table*}[htpb]
\caption{Summary of survey results}
\label{table:SummarySurvey} \centering
\smallskip

\begin{tabular}
{m{4.5cm}m{1.1cm}m{1.1cm}m{1.1cm}m{1.2cm}m{1.2cm}m{1.2cm}||m{2.6cm}|}
\cline{8-8}
& & & & & &  & \vtop{\hbox{\strut Respond. confidence}\hbox{\strut degree (1-4)}} \\
\end{tabular}

\begin{tabular}
{|m{4.5cm}|>{\centering\arraybackslash}m{1.1cm}|>{\centering\arraybackslash}m{1.1cm}|>{\centering\arraybackslash}m{1.1cm}|>{\centering\arraybackslash}m{1.2cm}|>{\centering\arraybackslash}m{1.1cm}||>{\centering\arraybackslash}m{1.1cm}||>{\centering\arraybackslash}m{1.1cm}|>{\centering\arraybackslash}m{1.1cm}|}
\cline{8-9}
\hline
\backslashbox{Question(finding)}{Answer} & \textbf{Strong agreement}  & \textbf{Agree-ment}  & \textbf{Weak agreement}  & \textbf{Weak disagreement}  & \textbf{Disagree-ment} & \textbf{\# of answers} & \textbf{Average} & \textbf{Std. deviation}\\
 \hline
The most frequently used CS detection techniques are based on rule-based approaches (F1) & 35.3\% & \cellcolor[HTML]{EFEFEF}47.1\% & 11.8\% & 5.9\% & 0.0\% & 34 & 3.2 & 0.8\\
\hline
Very few CS detection studies provide their oracles (a tagged dataset for training detection algorithms) (F2) & 26.5\% & \cellcolor[HTML]{EFEFEF}58.8\% & 11.8\% & 2.9\% & 0.0\% & 34 & 3.1 & 0.7 \\ \hline
In the detection of simpler CS (e.g. \textit{Long Method} or \textit{God Class}), the achieved precision and recall of detection techniques can be very high (up to 100\%) (F6) & 11.8\% & \cellcolor[HTML]{EFEFEF}44.1\% & 26.5\% & 0.0\% & 14.7\% & 34 & 3.2 & 0.5\\ \hline
When the complexity of CS is greater (e.g. \textit{Divergent Change} or \textit{Shotgun Surgery}), the precision and recall in detection are much lower than in simpler CS (F6) & 11.8\% & \cellcolor[HTML]{EFEFEF}47.1\% & 26.5\% & 8.8\% & 5.9\% & 34 & 3.1 & 0.7\\ \hline
There are few oracles (a tagged dataset for training detection algorithms) shared and publicly available. The existence of shared and collaborative oracles could improve the state of the art in CS detection research (F2)& \cellcolor[HTML]{EFEFEF}60.0\% & 34.3\% & 2.9\% & 2.9\% & 0.0\% & 35 & 3.6 & 0.5 \\ \hline
The vast majority of CS detection studies do not propose visualization features for their detection (F11) & 15.4\% & \cellcolor[HTML]{EFEFEF}66.7\% & 10.3\% & 5.1\% & 2.6\% & 39 & 3.0 & 1.0 \\ \hline
The vast majority of existing CS visualization studies did not present evidence of its usage upon large software systems (F11) & 12.5\% & \cellcolor[HTML]{EFEFEF}43.8\% & 34.4\% & 6.3\% & 0.0\% & 32 & 2.9 & 0.9 \\ \hline
Software visualization researchers have not adopted specific visualization related taxonomies (F11) & 9.4\% & 28.1\% & \cellcolor[HTML]{EFEFEF}46.9\% & 9.4\% & 6.3\% & 32 & 2.0 & 1.2 \\ \hline
If visualization related taxonomies were used in the implementation of CS detection tools, that could enhance their effectiveness (F11) & 11.8\% & \cellcolor[HTML]{EFEFEF}38.2\% & \cellcolor[HTML]{EFEFEF}38.2\% & 5.9\% & 5.9\% & 34 & 2.8 & 1.1 \\ \hline
The combined use of collaboration (among software developers) and visual resources may increase the effectiveness of CS detection (F11) & 23.5\% & \cellcolor[HTML]{EFEFEF}50.0\% & 26.5\% & 0.0\% & 0.0\% & 34 & 3.2 & 0.8 \\ \hline
\end{tabular}
\end{table*}

\subsection{Validity threats}
\label{sec:validitythreats}
We now go through the types of validity threats and corresponding mitigating actions that were considered in this study.

\textit{Conclusion validity.} We defined a data extraction form to ensure consistent extraction of relevant data for answering the research questions, therefore avoiding bias. The findings and implications are based on the extracted data.

\textit{Internal validity.} To avoid bias during the selection of studies to be included in this review, we used a thorough selection process, comprised of multiple stages. To reduce the possibility of missing relevant studies, in addition to the automatic search, we also used snowballing for complementary search.

\textit{External validity.} We have selected studies on code smells detection and visualization. The exclusion of studies on related subjects (e.g. refactoring and technical debt) may have caused some studies also dealing with code smells detection and visualization not to be included. However, we have found this situation to occur in breadth papers (covering a wide range of topics) rather than in depth (covering a specific topic). Since the latter are the more important ones for primary studies selection, we are confident on the relevance of the selected sample.

\textit{Construct validity.} The studies identified from the systematic review were accumulated from multiple literature databases covering relevant journals and proceedings. In the selection process the first author made the first selection and the remaining ones verified and confirmed it. To avoid bias in the selection of publications we specified and used a research protocol including the research questions and objectives of the study, inclusion and exclusion criteria, quality criteria, search strings, and strategy for search and for data extraction.

\section{Conclusion}
\label{sec:conclusion}

\subsection{Conclusions on this SLR}
\label{sec:SLRconclusions}
This paper presents a Systematic Literature Review with a twofold goal: the first is to identify the main CS detection techniques, and their effectiveness, as discussed in the literature, and the second is to analyze to which extent visual techniques have been applied to support practitioners in daily activities related to CS. For this purpose, we have specified 3 research questions (RQ1 through RQ3).

We applied our search string in six repositories (ACM Digital Library, IEEE Xplore, ISI Web of Science, Science Direct, Scopus, Springer Link) and complemented it with a manual search (backward snowballing), having obtained 1883 papers in total. After removing the duplicates, applying the inclusion and exclusion criteria, and quality criteria, we obtained 83 studies to analyze. Most of the studies were published in conference proceedings (76\%), followed by journals (23\%), and books (1\%). 
The 83 studies were analysed on the basis of 11 points (findings) related to the approach used for CS detection, dataset availability, programming languages supported, CS detected, evaluation of techniques, tools created, thresholds, validation and replication, and use of visualization techniques.

Regarding RQ1, we conclude that the most frequently used detection techniques are based on search-based approaches, which mainly apply ML algorithms, followed by metric-based approaches. Very few studies provide the oracles used and most of them target open-source Java projects. The most commonly detected CS are \textit{God Class}, \textit{Feature Envy} and \textit{Long Method}, by this order. On average, each study detects 3 CS, but the most frequent case is detecting only 1 CS.

As for RQ2, in the detection of simpler CS (e.g. \textit{God Class}) 4 approaches are used (probabilistic, metric-based, symptom-based, and search-based) and authors claim to achieve 100\% precision and recall results. However, when the complexity of CS is greater, the results have much lower relevance and very few studies use them. Thus, the detection problem is very far to be solved, depending on the detection results of the CS used, of the software systems in which they are detected, of the threshold and oracle values.

Regarding RQ3, we found that most studies that detect CS do not put forward a corresponding visualization feature. Several visualization approaches have been proposed for representing the structure of programs, either in 2D (e.g. graph-based, polymetric views) or in 3D (e.g. city metaphors), where the objective of allowing to identify potentially harmful design issues is claimed. However, we only found three studies that proposed dedicated solutions for CS visualization.

\subsection{Open issues}
\label{sec:openissues}

Detecting and visualizing CS are nontrivial endeavors. While producing this SLR we obtained a comprehensive perspective on the past and ongoing research in those fields, that allowed the identification of several open research issues. We briefly overview each of those issues, in the expectation it may inspire new researchers in the field.  

(1) Code smells subjective definitions hamper a shared interpretation across researchers' and practitioners' communities, thus hampering the advancement of the state-of-the-art and state-of-the-practice; to mitigate this problem it has been suggested a formal definition of CS (see \cite{Rasool2015}); a standardization effort, supported by an IT standards body, would certainly be a major initiative in this context;

(2) Open-source CS detection tooling is poor, both in language coverage (Java is dominant), and in CS coverage (e.g. only a small percentage of Fowler's catalog is supported);

(3) Primary studies reporting experiments on CS often do not make the corresponding scientific workflows and datasets available, thus not allowing their ``reproduction", where the goal is showing the correctness or validity of the published results;

(4) Replication of CS experiments, used to gain confidence in empirical findings, is also limited due to the effort of setting up the tooling required to running families of experiments, even when curated datasets on CS exist;

(5) Thresholds for deciding on CS occurrence are often arbitrary/unsubstantiated and not generalizable; in mitigation, we foresee the potential for the application of multi-criteria approaches that take into account the scope and context of CS, as well as approaches that explore the power of the crowd, such as the one proposed in \cite{Reis2017};

(6) CS studies in mobile and web environments are still scarce; due to their importance of those environments in nowadays life, we see a wide berth for CS research in those areas; 

(7) CS visualization techniques seem to have great potential, especially in large systems, to help developers in deciding if they agree with a CS occurrence suggested by an existing oracle; a large research effort is required to enlarge CS visualization diversity, both in scope (single method, single class, multiple classes) and coverage, since the existing literature only tackles a small percentage of the cataloged CS.

\begin{comment}
As future work, we plan the perception of CS and its use has been developed over time, answering questions such as: is there currently a greater attention to the detection of CS? Developers are more aware of the problem of CS? Is there any evolution in reducing CS subjectivity?
We also plan to develop a new approach to CS detection through the use of collective intelligence, which we call CrowdSmelling and presented in \cite{Reis2017}.
Finally, we plan to perform the forward snowballing technique by checking references of the selected studies to extend the number of relevant studies related to the research questions, as well as continue to update this SLR.
\end{comment}

\begin{acknowledgements}
This work was partially funded by the Portuguese Foundation for Science and Technology, under ISTAR's projects UIDB/ 04466/2020 and UIDP/04466/2020.
\end{acknowledgements}

%\section*{References}

\bibliographystyle{spbasic}      % basic style, author-year citations

\bibliography{main}

\pagebreak
\onecolumn

\appendix
\section*{\large{Appendices}}
\addcontentsline{toc}{section}{Appendices}
\renewcommand{\thesubsection}{\large{Appendix \Alph{subsection}.}}

\subsection{\large{Studies included in the review}}
\label{AppendixA}

\footnotesize{
%\begin{table}[htpb]
%\caption{Studies included in the review}
%\label{table:Studies} \centering
\begin{longtable}{|c|p{4.5cm}|p{3.5cm}|c|c|p{5cm}|} 
 \hline
 \textbf{ID} & \textbf{Title} & \textbf{Authors} & \textbf{Year} & \textbf{Publish type} & \textbf{Source title} \\\hline 
S1 & Java quality assurance by detecting code smells & E. van Emden; L. Moonen & 2002 & Conference & 9th Working Conference on Reverse Engineering (WCRE)\\ \hline 
S2 & Insights into system-wide code duplication & Rieger, M., Ducasse, S., Lanza, M. & 2004 & Conference & Working Conference on Reverse Engineering, IEEE Computer Society Press\\ \hline 
S3 & Detection strategies: Metrics-based rules for detecting design flaws & R. Marinescu & 2004 & Conference & 20th International Conference on Software Maintenance (ICSM). IEEE Computer Society Press \\ \hline 
S4 & Product metrics for automatic identification of "bad smell" design problems in Java source-code & M. J. Munro & 2005 & Conference & 11th IEEE International Software Metrics Symposium (METRICS'05) \\ \hline 
S5 & Multi-criteria detection of bad smells in code with UTA method & Walter B., Pietrzak B. & 2005 & Conference & International Conference on Extreme Programming and Agile Processes in Software Engineering (XP)\\ \hline

S6 & Adaptive detection of design flaws & Kreimer J. & 2005 & Conference & Fifth Workshop on Language Descriptions, Tools, and Applications (LDTA) \\ \hline

S7 & Visualization-Based Analysis of Quality for Large-Scale Software Systems & G. Langelier, H.A. Sahraoui,; P. Poulin & 2005 & Conference & 20th International Conference on Automated Software Engineering (ASE)\\ \hline 
S8 & Automatic generation of detection algorithms for design defects & Moha N., Guéhéneuc Y.-G., Leduc P. & 2006 & Conference & 21st IEEE/ACM International Conference on Automated Software Engineering (ASE)\\ \hline 
S9 & Object - Oriented Metrics in Practice & M. Lanza; R. Marinescu & 2006 & Book & Springer-Verlag\\ \hline 
S10 & Detecting Object Usage Anomalies & Andrzej Wasylkowski; Andreas Zeller; Christian Lindig & 2007 & Conference & 6th Joint Meeting of the European Software Engineering Conference and the ACM SIGSOFT Symposium on the Foundations of Software Engineering (ESEC/FSE)\\ \hline 
S11 & Empirically evaluating the usefulness of software visualization techniques in program comprehension activities & De F. Carneiro G., Orrico A.C.A., De Mendonça Neto M.G. & 2007 & Conference & VI Jornadas Iberoamericanas de Ingenieria de Software e Ingenieria del Conocimiento (JIISIC)\\ \hline 
S12 & A Catalogue of Lightweight Visualizations to Support Code Smell Inspection & Chris Parnin; Carsten Gorg; Ogechi Nnadi & 2008 & Conference & 4th ACM Symposium on Software Visualization (SoftVis)\\ \hline 
S13 & A domain analysis to specify design defects and generate detection algorithms & Moha N., Guéhéneuc Y.-G., Le Meur A.-F., Duchien L. & 2008 & Conference & International Conference on Fundamental Approaches to Software Engineering (FASE)\\ \hline 
S14 & JDeodorant: Identification and removal of type-checking bad smells & Tsantalis N., Chaikalis T., Chatzigeorgiou A. & 2008 & Conference & European Conference on Software Maintenance and Reengineering (CSMR) \\ \hline 
S15 & Empirical evaluation of clone detection using syntax suffix trees & Raimar Falk, Pierre Frenzel, Rainer Koschke & 2008 & Journal & Empirical Software Engineering \\ \hline 
S16 & Visual Detection of Design Anomalies & K. Dhambri, H. Sahraoui,; P. Poulin & 2008 & Conference & 12th European Conference on Software Maintenance and Reengineering (CSMR)\\ \hline

S17 & Visually localizing design problems with disharmony maps & Richard Wettel; Michele Lanza & 2008 & Conference & 4th ACM Symposium on Software Visualization (SoftVis)\\ \hline 
S18 & A Bayesian Approach for the Detection of Code and Design Smells & F. Khomh; S. Vaucher; Y. G. Gueheneuc; H. Sahraoui & 2009 & Conference & 9th International Conference on Quality Software (QSIC) \\ \hline

S19 & An Interactive Ambient Visualization for Code Smells & Emerson Murphy-Hill; Andrew P. Black & 2010 & Conference & 5th International Symposium on Software Visualization (SoftVis)\\ \hline

S20 & Learning from 6,000 Projects: Lightweight Cross-project Anomaly Detection & Natalie Gruska; Andrzej Wasylkowski; Andreas Zeller & 2010 & Conference & 19th International Symposium on Software Testing and Analysis \\ \hline

S21 & Identifying Code Smells with Multiple Concern Views & G. d. F. Carneiro; M. Silva; L. Mara; E. Figueiredo; C. Sant'Anna; A. Garcia; M. Mendonca & 2010 & Conference & Brazilian Symposium on Software Engineering (SBES)\\ \hline

S22 & Reducing Subjectivity in Code Smells Detection: Experimenting with the Long Method & S. Bryton; F. Brito e Abreu; M. Monteiro & 2010 & Conference & 7th International Conference on the Quality of Information and Communications Technology (QUATIC)\\ \hline 
S23 & DECOR: A method for the specification and detection of code and design smells & Moha N., Guéhéneuc Y.-G., Duchien L., Le Meur A.-F. & 2010 & Journal & IEEE Transactions on Software Engineering \\ \hline

S24 & IDS: An immune-inspired approach for the detection of software design smells & Hassaine S., Khomh F., Guéhéneucy Y.-G., Hamel S. & 2010 & Conference & 7th International Conference on the Quality of Information and Communications Technology (QUATIC)\\ \hline 

S25 & Detecting Missing Method Calls in Object-Oriented Software & Martin Monperrus Marcel Bruch Mira Mezini & 2010 & Conference & European Conference on Object-Oriented Programming (ECOOP)\\ \hline

S26 & From a domain analysis to the specification and detection of code and design smells & Naouel Moha, Yann-Gaël Guéhéneuc, Anne-Françoise Le Meur, Laurence Duchien, Alban Tiberghien & 2010 & Journal & Formal Aspects of Computing \\ \hline

S27 & BDTEX: A GQM-based Bayesian approach for the detection of antipatterns & Khomh F., Vaucher S., Guéhéneuc Y.-G., Sahraoui H. & 2011 & Journal & Journal of Systems and Software \\ \hline

S28 & IDE-based Real-time Focused Search for Near-miss Clones & Minhaz F. Zibran; Chanchal K. Roy & 2012 & Conference & 27th Annual ACM Symposium on Applied Computing (SAC) \\ \hline

S29 & Detecting Bad Smells with Weight Based Distance Metrics Theory & J. Dexun; M. Peijun; S. Xiaohong; W. Tiantian & 2012 & Conference & 2nd International Conference on Instrumentation, Measurement, Computer, Communication and Control (IMCCC)\\ \hline

S30 & Analytical learning based on a meta-programming approach for the detection of object-oriented design defects & Mekruksavanich S., Yupapin P.P., Muenchaisri P. & 2012 & Journal & Information Technology Journal \\ \hline

S31 & A New Design Defects Classification: Marrying Detection and Correction & Rim Mahouachi, Marouane Kessentini, Khaled Ghedira & 2012 & Conference & Fundamental Approaches to Software Engineering \\ \hline

S32 & Clones in Logic Programs and How to Detect Them & Céline Dandois, Wim Vanhoof & 2012 & Conference & Logic-Based Program Synthesis and Transformation \\ \hline

S33 & Smurf: A svm-based incremental anti-pattern detection approach & Maiga, A., Ali, N., Bhattacharya, N., Sabane, A., Guéhéneuc, Y-G, \& Aimeur, E & 2012 & Conference & 19th Working Conference on Reverse Engineering (WCRE)\\ \hline

S34 & Support vector machines for anti- pattern detection & Maiga A, Ali N, Bhattacharya N, Sabané A, Guéhéneuc Y-G, Antoniol G, Aïmeur E & 2012 & Conference & 27th IEEE/ACM International Conference on Automated Software Engineering (ASE)\\ \hline

S35 & Detecting Missing Method Calls As Violations of the Majority Rule & Martin Monperrus; Mira Mezini & 2013 & Journal & ACM Transactions on Software Engineering Methodology \\ \hline

S36 & Code Smell Detection: Towards a Machine Learning-Based Approach & F. A. Fontana; M. Zanoni; A. Marino; M. V. Mantyla; & 2013 & Conference & 29th IEEE International Conference on Software Maintenance (ICSM)\\ \hline

S37 & Identification of Refused Bequest Code Smells & E. Ligu; A. Chatzigeorgiou; T. Chaikalis; N. Ygeionomakis & 2013 & Conference & 29th IEEE International Conference on Software Maintenance (ICSM)\\ \hline

S38 & JSNOSE: Detecting JavaScript Code Smells & A. M. Fard; A. Mesbah & 2013 & Conference & 13th International Working Conference on Source Code Analysis and Manipulation (SCAM)\\ \hline

S39 & Interactive ambient visualizations for soft advice & Murphy-Hill E., Barik T., Black A.P. & 2013 & Journal & Information Visualization \\ \hline

S40 & A novel approach to effective detection and analysis of code clones & Rajakumari K.E., Jebarajan T. & 2013 & Conference & 3rd International Conference on Innovative Computing Technology (INTECH)\\ \hline

S41 & Competitive coevolutionary code-smells detection & Boussaa M., Kessentini W., Kessentini M., Bechikh S., Ben Chikha S. & 2013 & Conference & International Symposium on Search Based Software Engineering (SSBSE)\\ \hline

S42 & Detecting bad smells in source code using change history information & Palomba F., Bavota G., Di Penta M., Oliveto R., De Lucia A., Poshyvanyk D. & 2013 & Conference & 28th International Conference on Automated Software Engineering (ASE). IEEE/ACM\\ \hline

S43 & Code-Smell Detection As a Bilevel Problem & Dilan Sahin; Marouane Kessentini; Slim Bechikh; Kalyanmoy Deb & 2014 & Journal & ACM Transactions on Software Engineering Methodology\\ \hline

S44 & Two level dynamic approach for Feature Envy detection & S. Kumar; J. K. Chhabra & 2014 & Conference & International Conference on Computer and Communication Technology (ICCCT).\\ \hline

S45 & A Cooperative Parallel Search-Based Software Engineering Approach for Code-Smells Detection & Kessentini W., Kessentini M., Sahraoui H., Bechikh S., Ouni A. & 2014 & Journal & IEEE Transactions on Software Engineering \\ \hline

S46 & SourceMiner: Towards an Extensible Multi-perspective Software Visualization Environment & Glauco de Figueiredo Carneiro, Manoel Gomes de Mendonça Neto & 2014 & Conference & International Conference on Enterprise Information Systems (ICEIS)\\ \hline

S47 & Including Structural Factors into the Metrics-based Code Smells Detection & Bartosz Walter; Błażej Matuszyk; Francesca Arcelli Fontana & 2015 & Conference & XP'2015 Workshops\\ \hline

S48 & Textual Analysis for Code Smell Detection & Fabio Palomba & 2015 & Conference & 37th International Conference on Software Engineering\\ \hline

S49 & Using Developers' Feedback to Improve Code Smell Detection & Mario Hozano; Henrique Ferreira; Italo Silva; Baldoino Fonseca; Evandro Costa & 2015 & Conference & 30th Annual ACM Symposium on Applied Computing (SAC)\\ \hline

S50 & Code Bad Smell Detection through Evolutionary Data Mining & S. Fu; B. Shen & 2015 & Conference & 2015 ACM/IEEE International Symposium on Empirical Software Engineering and Measurement (ESEM) \\ \hline

S51 & Mining Version Histories for Detecting Code Smells & F. Palomba; G. Bavota; M. D. Penta; R. Oliveto; D. Poshyvanyk; A. De Lucia & 2015 & Conference & IEEE Transactions on Software Engineering\\ \hline

S52 & Detection and handling of model smells for MATLAB/simulink models & Gerlitz T., Tran Q.M., Dziobek C. & 2015 & Conference & CEUR Workshop Proceedings\\ \hline

S53 & Experience report: Evaluating the effectiveness of decision trees for detecting code smells & Amorim L., Costa E., Antunes N., Fonseca B., Ribeiro M. & 2015 & Conference & 26th International Symposium on Software Reliability Engineering (ISSRE)\\ \hline

S54 & Detecting software design defects using relational association rule mining & Gabriela Czibula, Zsuzsanna Marian, Istvan Gergely Czibula & 2015 & Journal & Knowledge and Information Systems\\ \hline

S55 & A Graph-based Approach to Detect Unreachable Methods in Java Software & Simone Romano; Giuseppe Scanniello; Carlo Sartiani; Michele Risi & 2016 & Conference & 31st Annual ACM Symposium on Applied Computing (SAC)\\ \hline

S56 & Comparing and experimenting machine learning techniques for code smell detection & Arcelli Fontana F., Mäntylä M.V., Zanoni M., Marino A. & 2016 & Journal & Empirical Software Engineering \\ \hline

S57 & A Lightweight Approach for Detection of Code Smells & Ghulam Rasool, Zeeshan Arshad & 2016 & Journal & Arabian Journal for Science and Engineering \\ \hline

S58 & Multi-objective code-smells detection using good and bad design examples & Usman Mansoor, Marouane Kessentini, Bruce R. Maxim, Kalyanmoy Deb & 2016 & Journal & Software Quality Journal \\ \hline

S59 & Continuous Detection of Design Flaws in Evolving Object-oriented Programs Using Incremental Multi-pattern Matching & Sven Peldszus; Géza Kulcsár; Malte Lochau; Sandro Schulze & 2016 & Conference & 31st IEEE/ACM International Conference on Automated Software Engineering (ASE)\\ \hline

S60 & Metric and rule based automated detection of antipatterns in object-oriented software systems & M. T. Aras, Y. E. Selçuk & 2016 & Conference & 2016 7th International Conference on Computer Science and Information Technology (CSIT) \\ \hline

S61 & Automated detection of code smells caused by null checking conditions in Java programs & K. Sirikul, C. Soomlek & 2016 & Conference & 2016 13th International Joint Conference on Computer Science and Software Engineering (JCSSE) \\ \hline

S62 & A textual-based technique for Smell Detection & F. Palomba, A. Panichella, A. De Lucia, R. Oliveto, A. Zaidman & 2016 & Conference & 24th International Conference on Program Comprehension (ICPC)\\ \hline

S63 & Detecting Code Smells in Python Programs & Z. Chen, L. Chen, W. Ma, B. Xu & 2016 & Conference & 2016 International Conference on Software Analysis; Testing and Evolution (SATE)\\ \hline

S64 & Interactive Code Smells Detection: An Initial Investigation & Mkaouer, Mohamed Wiem & 2016 & Conference & Symposium on Search-Based Software Engineering (SSBSE) \\ \hline

S65 & Detecting shotgun surgery bad smell using similarity measure distribution model & Saranya G., Khanna Nehemiah H., Kannan A., Vimala S. & 2016 & Journal & Asian Journal of Information Technology \\ \hline

S66 & Detecting Android Smells Using Multi-objective Genetic Programming & Marouane Kessentini; Ali Ouni & 2017 & Conference & 4th International Conference on Mobile Software Engineering and Systems (MOBILESoft) \\ \hline

S67 & Smells Are Sensitive to Developers!: On the Efficiency of (Un)Guided Customized Detection & Mario Hozano; Alessandro Garcia; Nuno Antunes; Baldoino Fonseca; Evandro Costa & 2017 & Conference & 25th International Conference on Program Comprehension (ICPC)\\ \hline

S68 & An automated code smell and anti-pattern detection approach & S. Velioglu, Y. E. Selçuk & 2017 & Conference & 2017 IEEE 15th International Conference on Software Engineering Research; Management and Applications (SERA) \\ \hline

S69 & Lightweight detection of Android-specific code smells: The aDoctor project & Palomba F., Di Nucci D., Panichella A., Zaidman A., De Lucia A. & 2017 & Conference & 24th IEEE International Conference on Software Analysis, Evolution, and Reengineering (SANER)\\ \hline

S70 & On the Use of Smelly Examples to Detect Code Smells in JavaScript & Ian Shoenberger, Mohamed Wiem Mkaouer, Marouane Kessentini & 2017 & Conference & European Conference on the Applications of Evolutionary Computation (EvoApplications)\\ \hline

S71 & A Support Vector Machine Based Approach for Code Smell Detection & A. Kaur; S. Jain; S. Goel & 2017 & Conference & International Conference on Machine Learning and Data Science (MLDS) \\ \hline

S72 & c-JRefRec: Change-based identification of Move Method refactoring opportunities & N. Ujihara; A. Ouni; T. Ishio; K. Inoue & 2017 & Conference & 24th International Conference on Software Analysis, Evolution and Reengineering (SANER) \\ \hline

S73 & A Feature Envy Detection Method Based on Dataflow Analysis & W. Chen; C. Liu; B. Li & 2018 & Conference & 42nd Annual Computer Software and Applications Conference (COMPSAC) \\ \hline

S74 & A Hybrid Approach To Detect Code Smells using Deep Learning & Hadj-Kacem, M; Bouassida, N & 2018 & Conference & 13th International Conference on Evaluation of Novel Approaches to Software Engineering (ENASE)\\ \hline

S75 & Deep Learning Based Feature Envy Detection & Hui Liu and Zhifeng Xu and Yanzhen Zou & 2018 & Conference & 33rd ACM/IEEE International Conference on Automated Software Engineering (ASE)\\ \hline

S76 & Detecting Bad Smells in Software Systems with Linked Multivariate Visualizations & H. Mumtaz; F. Beck; D. Weiskopf & 2018 & Conference & Working Conference on Software Visualization (VisSoft) \\ \hline

S77 & Detecting code smells using machine learning techniques: Are we there yet? & D. Di Nucci; F. Palomba; D. A. Tamburri; A. Serebrenik; A. De Lucia & 2018 & Conference & 25th International Conference on Software Analysis, Evolution and Reengineering (SANER) \\ \hline

S78 & Exploring the Use of Rapid Type Analysis for Detecting the Dead Method Smell in Java Code & S. Romano; G. Scanniello & 2018 & Conference & 44th Euromicro Conference on Software Engineering and Advanced Applications (SEAA) \\ \hline

S79 & Model level code smell detection using EGAPSO based on similarity measures & Saranya, G; Nehemiah, HK; Kannan, A; Nithya, V & 2018 & Journal & Alexandria Engineering Journal\\ \hline

S80 & Software Code Smell Prediction Model Using Shannon, Renyi and Tsallis Entropies & Gupta, A; Suri, B; Kumar, V; Misra, S; Blazauskas, T; Damasevicius, R & 2018 & Journal & Entropy \\ \hline

S81 & Towards Feature Envy Design Flaw Detection at Block Level & Ã. Kiss; P. F. Mihancea & 2018 & Conference & International Conference on Software Maintenance and Evolution (ICSME) \\ \hline

S82 & Understanding metric-based detectable smells in Python software: A comparative study & Chen, ZF; Chen, L; Ma, WWY; Zhou, XY; Zhou, YM; Xu, BW & 2018 & Journal & Information and Software Technology\\ \hline

S83 & SP-J48: a novel optimization and machine-learning-based approach for solving complex problems: special application in software engineering for detecting code smells & Amandeep Kaur, Sushma Jain, Shivani Goel & 2019 & Journal & Neural Computing and Applications \\ \hline

\end{longtable}
%\end{table}
}
\pagebreak

\subsection{\large{Studies after applying inclusion and exclusion criteria (phase 3)}}
\label{AppendixB}

\footnotesize{
%\begin{table}[htpb]
%\caption{Studies included in the review}
%\label{table:Studies} \centering
\begin{longtable}{|c|p{4cm}|p{4cm}|c|c|p{5cm}|} 
 \hline
 \textbf{ID} & \textbf{Title} & \textbf{Authors} & \textbf{Year} & \textbf{Publish type} & \textbf{Source title} \\\hline 
1 & Java quality assurance by detecting code smells & E. van Emden; L. Moonen & 2002 & Conference & 9th Working Conference on Reverse Engineering (WCRE)\\ \hline

2 & Insights into system-wide code duplication & Rieger, M., Ducasse, S., Lanza, M. & 2004 & Conference & 11th Working Conference on Reverse Engineering (WCRE)\\ \hline

3 & Detection strategies: Metrics-based rules for detecting design flaws & R. Marinescu & 2004 & Conference & 20th International Conference on Software Maintenance (ICSM)\\ \hline

4 & Product metrics for automatic identification of "bad smell" design problems in Java source-code & M. J. Munro & 2005 & Conference & 11th IEEE International Software Metrics Symposium (METRICS'05) \\ \hline

5 & Multi-criteria detection of bad smells in code with UTA method & Walter B., Pietrzak B. & 2005 & Conference & International Conference on Extreme Programming and Agile Processes in Software Engineering (XP) \\ \hline

6 & Adaptive detection of design flaws & Kreimer J. & 2005 & Conference & Fifth Workshop on Language Descriptions, Tools, and Applications (LDTA)\\ \hline 

7 & Visualization-Based Analysis of Quality for Large-Scale Software Systems & G. Langelier, H.A. Sahraoui,; P. Poulin & 2005 & Conference & 20th International Conference on Automated Software Engineering (ASE)\\ \hline

8 & Automatic generation of detection algorithms for design defects & Moha N., Guéhéneuc Y.-G., Leduc P. & 2006 & Conference & 21st IEEE/ACM International Conference on Automated Software Engineering (ASE)\\ \hline

9 & Object - Oriented Metrics in Practice & M. Lanza; R. Marinescu & 2006 & Book & Springer-Verlag\\ \hline

10 & Detecting Object Usage Anomalies & Andrzej Wasylkowski; Andreas Zeller; Christian Lindig & 2007 & Conference & 6th Joint Meeting of the European Software Engineering Conference and the ACM SIGSOFT Symposium on the Foundations of Software Engineering (ESEC/FSE)\\ \hline

11 & Using Concept Analysis to Detect Co-change Patterns & Tudor Girba; Stephane Ducasse; Adrian Kuhn; Radu Marinescu; Ratiu Daniel & 2007 & Conference & 9th International Workshop on Principles of Software Evolution: In Conjunction with the 6th ESEC/FSE Joint Meeting \\ \hline

12 & Empirically evaluating the usefulness of software visualization techniques in program comprehension activities & De F. Carneiro G., Orrico A.C.A., De Mendonça Neto M.G. & 2007 & Conference & VI Jornadas Iberoamericanas de Ingenieria de Software e Ingenieria del Conocimiento (JIISIC)\\ \hline

13 & A Catalogue of Lightweight Visualizations to Support Code Smell Inspection & Chris Parnin; Carsten Gorg; Ogechi Nnadi & 2008 & Conference & 4th ACM Symposium on Software Visualization (SoftVis)\\ \hline

14 & A domain analysis to specify design defects and generate detection algorithms & Moha N., Guéhéneuc Y.-G., Le Meur A.-F., Duchien L. & 2008 & Conference & Lecture Notes in Computer Science (including subseries Lecture Notes in Artificial Intelligence and Lecture Notes in Bioinformatics) \\ \hline

15 & JDeodorant: Identification and removal of type-checking bad smells & Tsantalis N., Chaikalis T., Chatzigeorgiou A. & 2008 & Conference & European Conference on Software Maintenance and Reengineering (CSMR)\\ \hline

16 & A Survey about the Intent to Use Visual Defect Annotations for Software Models & Jörg Rech, Axel Spriestersbach & 2008 & Conference & Model Driven Architecture – Foundations and Applications \\ \hline

17 & Empirical evaluation of clone detection using syntax suffix trees & Raimar Falk, Pierre Frenzel, Rainer Koschke & 2008 & Journal & Empirical Software Engineering \\ \hline 
18 & Visual Detection of Design Anomalies & K. Dhambri, H. Sahraoui, P. Poulin & 2008 & Conference & 12th European Conference on Software Maintenance and Reengineering (CSMR)\\ \hline 
19 & Detecting bad smells in object oriented design using design change propagation probability matrix & A. Rao; K. Raddy & 2008 & Conference & International MultiConference of Engineers and Computer Scientists (IMECS)\\ \hline

20 & Visually localizing design problems with disharmony maps & Richard Wettel; Michele Lanza & 2008 & Conference & 4th ACM Symposium on Software visualization (SoftVis)\\ \hline 
21 & A Bayesian Approach for the Detection of Code and Design Smells & F. Khomh; S. Vaucher; Y. G. Gueheneuc; H. Sahraoui & 2009 & Conference & 2009 Ninth International Conference on Quality Software \\ \hline

22 & A Flexible Framework for Quality Assurance of Software Artefacts with Applications to Java, UML, and TTCN-3 Test Specifications & J. Nodler; H. Neukirchen; J. Grabowski & 2009 & Conference & 2009 International Conference on Software Testing Verification and Validation (ICST)\\ \hline

23 & An Interactive Ambient Visualization for Code Smells & Emerson Murphy-Hill; Andrew P. Black & 2010 & Conference & 5th International Symposium on Software Visualization (SoftVis)\\ \hline

24 & Learning from 6,000 Projects: Lightweight Cross-project Anomaly Detection & Natalie Gruska; Andrzej Wasylkowski; Andreas Zeller & 2010 & Conference & 19th International Symposium on Software Testing and Analysis (ISSTA) \\ \hline

25 & Identifying Code Smells with Multiple Concern Views & G. d. F. Carneiro; M. Silva; L. Mara; E. Figueiredo; C. Sant'Anna; A. Garcia; M. Mendonca & 2010 & Conference & Brazilian Symposium on Software Engineering (SBES) \\ \hline

26 & Reducing Subjectivity in Code Smells Detection: Experimenting with the Long Method & S. Bryton; F. Brito e Abreu; M. Monteiro & 2010 & Conference & 7th International Conference on the Quality of Information and Communications Technology (QUATIC)\\ \hline

27 & DECOR: A method for the specification and detection of code and design smells & Moha N., Guéhéneuc Y.-G., Duchien L., Le Meur A.-F. & 2010 & Journal & IEEE Transactions on Software Engineering \\ \hline

28 & IDS: An immune-inspired approach for the detection of software design smells & Hassaine S., Khomh F., Guéhéneucy Y.-G., Hamel S. & 2010 & Conference & 7th International Conference on the Quality of Information and Communications Technology (QUATIC)\\ \hline

29 & Detecting Missing Method Calls in Object-Oriented Software & Martin Monperrus Marcel Bruch Mira Mezini & 2010 & Conference & European Conference on Object-Oriented Programming (ECOOP) \\ \hline

30 & From a domain analysis to the specification and detection of code and design smells & Naouel Moha, Yann-Gaël Guéhéneuc, Anne-Françoise Le Meur, Laurence Duchien, Alban Tiberghien & 2010 & Journal & Formal Aspects of Computing \\ \hline 

31 & BDTEX: A GQM-based Bayesian approach for the detection of antipatterns & Khomh F., Vaucher S., Guéhéneuc Y.-G., Sahraoui H. & 2011 & Journal & Journal of Systems and Software \\ \hline

32 & An Approach for Source Code Classification Using Software Metrics and Fuzzy Logic to Improve Code Quality with Refactoring Techniques & Pornchai Lerthathairat, Nakornthip Prompoon & 2011 & Conference & 2nd International Conference on Software Engineering and Computer Systems (ICSECS)\\ \hline

33 & IDE-based Real-time Focused Search for Near-miss Clones & Minhaz F. Zibran; Chanchal K. Roy & 2012 & Conference & 27th Annual ACM Symposium on Applied Computing (SAC)\\ \hline 
34 & Detecting Bad Smells with Weight Based Distance Metrics Theory & J. Dexun; M. Peijun; S. Xiaohong; W. Tiantian & 2012 & Conference & Second International Conference on Instrumentation, Measurement, Computer, Communication and Control (IMCCC) \\ \hline

35 & Analytical learning based on a meta-programming approach for the detection of object-oriented design defects & Mekruksavanich S., Yupapin P.P., Muenchaisri P. & 2012 & Journal & Information Technology Journal \\ \hline

36 & Automatic identification of the anti-patterns using the rule-based approach & Polášek I., Snopko S., Kapustík I. & 2012 & Conference & 10th Jubilee International Symposium on Intelligent Systems and Informatics (SISY) \\ \hline

37 & A New Design Defects Classification: Marrying Detection and Correction & Rim Mahouachi, Marouane Kessentini, Khaled Ghedira & 2012 & Conference &  International Conference on Fundamental Approaches to Software Engineering (FASE)\\ \hline

38 & Clones in Logic Programs and How to Detect Them & Céline Dandois, Wim Vanhoof & 2012 & Conference & International Symposium on Logic-Based Program Synthesis and Transformation (LOPSTR)\\ \hline

39 & Smurf: A svm-based incremental anti-pattern detection approach & Maiga, A., Ali, N., Bhattacharya, N., Sabane, A., Guéhéneuc, Y. G., \& Aimeur, E & 2012 & Conference & 19th Working Conference on Reverse Engineering (WCRE)\\ \hline

40 & Support vector machines for anti- pattern detection & Maiga A, Ali N, Bhattacharya N, Sabané A, Guéhéneuc Y-G, Antoniol G, Aïmeur E & 2012 & Conference & 27th IEEE/ACM International Conference on Automated Software Engineering (ASE)\\ \hline

41 & Detecting Missing Method Calls As Violations of the Majority Rule & Martin Monperrus; Mira Mezini & 2013 & Journal & ACM Transactions on Software Engineering Methodology \\ \hline

42 & Code Smell Detection: Towards a Machine Learning-Based Approach & F. A. Fontana; M. Zanoni; A. Marino; M. V. Mantyla; & 2013 & Conference & 29th IEEE International Conference on Software Maintenance (ICSM)\\ \hline

43 & Identification of Refused Bequest Code Smells & E. Ligu; A. Chatzigeorgiou; T. Chaikalis; N. Ygeionomakis & 2013 & Conference & 29th IEEE International Conference on Software Maintenance (ICSM)\\ \hline

44 & JSNOSE: Detecting JavaScript Code Smells & A. M. Fard; A. Mesbah & 2013 & Conference & 13th International Working Conference on Source Code Analysis and Manipulation (SCAM)\\ \hline

45 & Interactive ambient visualizations for soft advice & Murphy-Hill E., Barik T., Black A.P. & 2013 & Journal & Information Visualization \\ \hline

46 & A novel approach to effective detection and analysis of code clones & Rajakumari K.E., Jebarajan T. & 2013 & Conference & 3rd International Conference on Innovative Computing Technology (INTECH)\\ \hline 

47 & Competitive coevolutionary code-smells detection & Boussaa M., Kessentini W., Kessentini M., Bechikh S., Ben Chikha S. & 2013 & Conference & Lecture Notes in Computer Science (including subseries Lecture Notes in Artificial Intelligence and Lecture Notes in Bioinformatics) \\ \hline

48 & Detecting bad smells in source code using change history information & Palomba F., Bavota G., Di Penta M., Oliveto R., De Lucia A., Poshyvanyk D. & 2013 & Conference & 28th IEEE/ACM International Conference on Automated Software Engineering (ASE) \\ \hline

49 & Code-Smell Detection As a Bilevel Problem & Dilan Sahin; Marouane Kessentini; Slim Bechikh; Kalyanmoy Deb & 2014 & Journal & ACM Trans. Softw. Eng. Methodol. \\ \hline

50 & Two level dynamic approach for Feature Envy detection & S. Kumar; J. K. Chhabra & 2014 & Conference & International Conference on Computer and Communication Technology (ICCCT)\\ \hline

51 & A Cooperative Parallel Search-Based Software Engineering Approach for Code-Smells Detection & Kessentini W., Kessentini M., Sahraoui H., Bechikh S., Ouni A. & 2014 & Journal & IEEE Transactions on Software Engineering \\ \hline 

52 & SourceMiner: Towards an Extensible Multi-perspective Software Visualization Environment & Glauco de Figueiredo Carneiro, Manoel Gomes de Mendonça Neto & 2014 & Conference & International Conference on Enterprise Information Systems (ICEIS)\\ \hline 

53 & Including Structural Factors into the Metrics-based Code Smells Detection & Bartosz Walter; Błażej Matuszyk; Francesca Arcelli Fontana & 2015 & Conference & XP'2015 Workshops \\ \hline

54 & Textual Analysis for Code Smell Detection & Fabio Palomba & 2015 & Conference & 37th International Conference on Software Engineering (ICSE)\\ \hline 

55 & Using Developers' Feedback to Improve Code Smell Detection & Mario Hozano; Henrique Ferreira; Italo Silva; Baldoino Fonseca; Evandro Costa & 2015 & Conference & 30th Annual ACM Symposium on Applied Computing (SAC)\\ \hline

56 & Code Bad Smell Detection through Evolutionary Data Mining & S. Fu; B. Shen & 2015 & Conference & 2015 ACM/IEEE International Symposium on Empirical Software Engineering and Measurement (ESEM) \\ \hline 

57 & JSpIRIT: a flexible tool for the analysis of code smells & S. Vidal; H. Vazquez; J. A. Diaz-Pace; C. Marcos; A. Garcia; W. Oizumi & 2015 & Conference & 34th International Conference of the Chilean Computer Science Society (SCCC) \\ \hline

58 & Mining Version Histories for Detecting Code Smells & F. Palomba; G. Bavota; M. D. Penta; R. Oliveto; D. Poshyvanyk; A. De Lucia & 2015 & Conference & IEEE Transactions on Software Engineering\\ \hline

59 & Detection and handling of model smells for MATLAB/simulink models & Gerlitz T., Tran Q.M., Dziobek C. & 2015 & Conference & CEUR Workshop Proceedings \\ \hline

60 & Experience report: Evaluating the effectiveness of decision trees for detecting code smells & Amorim L., Costa E., Antunes N., Fonseca B., Ribeiro M. & 2015 & Conference & 26th International Symposium on Software Reliability Engineering (ISSRE)\\ \hline

61 & Detecting software design defects using relational association rule mining & Gabriela Czibula, Zsuzsanna Marian, Istvan Gergely Czibula & 2015 & Journal & Knowledge and Information Systems \\ \hline

62 & A Graph-based Approach to Detect Unreachable Methods in Java Software & Simone Romano; Giuseppe Scanniello; Carlo Sartiani; Michele Risi & 2016 & Conference & 31st Annual ACM Symposium on Applied Computing (SAC)\\ \hline

63 & Comparing and experimenting machine learning techniques for code smell detection & Arcelli Fontana F., Mäntylä M.V., Zanoni M., Marino A. & 2016 & Journal & Empirical Software Engineering \\ \hline

64 & A Lightweight Approach for Detection of Code Smells & Ghulam Rasool, Zeeshan Arshad & 2016 & Journal & Arabian Journal for Science and Engineering \\ \hline

65 & Multi-objective code-smells detection using good and bad design examples & Usman Mansoor, Marouane Kessentini, Bruce R. Maxim, Kalyanmoy Deb & 2016 & Journal & Software Quality Journal \\ \hline

66 & Continuous Detection of Design Flaws in Evolving Object-oriented Programs Using Incremental Multi-pattern Matching & Sven Peldszus; Géza Kulcsár; Malte Lochau; Sandro Schulze & 2016 & Conference & 31st IEEE/ACM International Conference on Automated Software Engineering (ASE)\\ \hline

67 & Metric and rule based automated detection of antipatterns in object-oriented software systems & M. T. Aras, Y. E. Selçuk & 2016 & Conference & 7th International Conference on Computer Science and Information Technology (CSIT) \\ \hline

68 & Automated detection of code smells caused by null checking conditions in Java programs & K. Sirikul, C. Soomlek & 2016 & Conference & 13th International Joint Conference on Computer Science and Software Engineering (JCSSE) \\ \hline

69 & A textual-based technique for Smell Detection & F. Palomba, A. Panichella, A. De Lucia, R. Oliveto, A. Zaidman & 2016 & Conference & 24th International Conference on Program Comprehension (ICPC) \\ \hline

70 & Detecting Code Smells in Python Programs & Z. Chen, L. Chen, W. Ma, B. Xu & 2016 & Conference & International Conference on Software Analysis, Testing and Evolution (SATE) \\ \hline

71 & DT : a detection tool to automatically detect code smell in software project & Liu, Xinghua; Zhang, Cheng & 2016 & Conference & 4th International Conference on Machinery, Materials and Information Technology Applications \\ \hline 

72 & Interactive Code Smells Detection: An Initial Investigation & Mkaouer, Mohamed Wiem & 2016 & Conference & Symposium on Search-Based Software Engineering (SSBSE)\\ \hline

73 & Automatic detection of bad smells from code changes & Hammad M., Labadi A. & 2016 & Journal & International Review on Computers and Software \\ \hline

74 & Detecting shotgun surgery bad smell using similarity measure distribution model & Saranya G., Khanna Nehemiah H., Kannan A., Vimala S. & 2016 & Journal & Asian Journal of Information Technology \\ \hline 

75 & Detecting Android Smells Using Multi-objective Genetic Programming & Marouane Kessentini; Ali Ouni & 2017 & Conference & 4th International Conference on Mobile Software Engineering and Systems (MOBILESoft)\\ \hline

76 & Smells Are Sensitive to Developers!: On the Efficiency of (Un)Guided Customized Detection & Mario Hozano; Alessandro Garcia; Nuno Antunes; Baldoino Fonseca; Evandro Costa & 2017 & Conference & 25th International Conference on Program Comprehension \\ \hline

77 & An arc-based approach for visualization of code smells & M. Steinbeck & 2017 & Conference & 24th International Conference on Software Analysis; Evolution and Reengineering (SANER). IEEE \\ \hline

78 & An automated code smell and anti-pattern detection approach & S. Velioglu, Y. E. Selçuk & 2017 & Conference & 15th International Conference on Software Engineering Research; Management and Applications (SERA) \\ \hline

79 & Lightweight detection of Android-specific code smells: The aDoctor project & Palomba F., Di Nucci D., Panichella A., Zaidman A., De Lucia A. & 2017 & Conference & 24th IEEE International Conference on Software Analysis, Evolution, and Reengineering (SANER)\\ \hline

80 & On the Use of Smelly Examples to Detect Code Smells in JavaScript & Ian Shoenberger, Mohamed Wiem Mkaouer, Marouane Kessentini & 2017 & Conference & European Conference on the Applications of Evolutionary Computation (EvoApplications)\\ \hline

81 & A Support Vector Machine Based Approach for Code Smell Detection & A. Kaur; S. Jain; S. Goel & 2017 & Conference & International Conference on Machine Learning and Data Science (MLDS) \\ \hline

82 & An ontology-based approach to analyzing the occurrence of code smells in software & Da Silva Carvalho, L.P., Novais, R., Do Nascimento Salvador, L., De Mendonça Neto, M.G. & 2017 & Conference & 19th International Conference on Enterprise Information Systems (ICEIS)\\ \hline

83 & Automatic multiprogramming bad smell detection with refactoring & Verma, A; Kumar, A; Kaur, I & 2017 & Journal & International Journal of Advanced and Applied Sciences\\ \hline 
84 & c-JRefRec: Change-based identification of Move Method refactoring opportunities & N. Ujihara; A. Ouni; T. Ishio; K. Inoue & 2017 & Conference & 24th International Conference on Software Analysis, Evolution and Reengineering (SANER) \\ \hline

85 & Finding bad code smells with neural network models & Kim, D.K. & 2017 & Journal & International Journal of Electrical and Computer Engineering \\ \hline

86 & Metric based detection of refused bequest code smell & B. M. Merzah; Y. E. SelÃ§uk & 2017 & Conference & 9th International Conference on Computational Intelligence and Communication Networks (CICN) \\ \hline

87 & Systematic exhortation of code smell detection using JSmell for Java source code & M. Sangeetha; P. Sengottuvelan & 2017 & Conference & International Conference on Inventive Systems and Control (ICISC) \\ \hline

88 & A Feature Envy Detection Method Based on Dataflow Analysis & W. Chen; C. Liu; B. Li & 2018 & Conference & 42nd Annual Computer Software and Applications Conference (COMPSAC) \\ \hline

89 & A Hybrid Approach To Detect Code Smells using Deep Learning & Hadj-Kacem, M; Bouassida, N & 2018 & Conference & 13th International Conference on Evaluation of Novel Approaches to Software Engineering (ENASE)\\ \hline

90 & Automatic detection of feature envy using machine learning techniques & Özkalkan, Z., Aydin, K., Tetik, H.Y., Sağlam, R.B. & 2018 & Conference & 12th Turkish National Software Engineering Symposium \\ \hline

91 & Code-smells identification by using PSO approach & Ramesh, G., Mallikarjuna Rao, C. & 2018 & Journal & International Journal of Recent Technology and Engineering \\ \hline

92 & Deep Learning Based Feature Envy Detection & Hui Liu and Zhifeng Xu and Yanzhen Zou & 2018 & Conference & 33rd ACM/IEEE International Conference on Automated Software Engineering (ASE)\\ \hline

93 & Detecting Bad Smells in Software Systems with Linked Multivariate Visualizations & H. Mumtaz; F. Beck; D. Weiskopf & 2018 & Conference & Working Conference on Software Visualization (VisSoft) \\ \hline

94 & Detecting code smells using machine learning techniques: Are we there yet? & D. Di Nucci; F. Palomba; D. A. Tamburri; A. Serebrenik; A. De Lucia & 2018 & Conference & 25th International Conference on Software Analysis, Evolution and Reengineering (SANER) \\ \hline

95 & DT: An Upgraded Detection Tool to Automatically Detect Two Kinds of Code Smell: Duplicated Code and Feature Envy & Xinghua Liu and Cheng Zhang & 2018 & Conference & International Conference on Geoinformatics and Data Analysis \\ \hline

96 & Exploring the Use of Rapid Type Analysis for Detecting the Dead Method Smell in Java Code & S. Romano; G. Scanniello & 2018 & Conference & 2018 44th Euromicro Conference on Software Engineering and Advanced Applications (SEAA) \\ \hline

97 & Model level code smell detection using EGAPSO based on similarity measures & Saranya, G; Nehemiah, HK; Kannan, A; Nithya, V & 2018 & Journal & Alexandria Engineering Journal\\ \hline 
98 & Software Code Smell Prediction Model Using Shannon, Renyi and Tsallis Entropies & Gupta, A; Suri, B; Kumar, V; Misra, S; Blazauskas, T; Damasevicius, R & 2018 & Journal & Entropy \\ \hline

99 & Towards Feature Envy Design Flaw Detection at Block Level & Ã. Kiss; P. F. Mihancea & 2018 & Conference & International Conference on Software Maintenance and Evolution (ICSME) \\ \hline 

100 & Understanding metric-based detectable smells in Python software: A comparative study & Chen, ZF; Chen, L; Ma, WWY; Zhou, XY; Zhou, YM; Xu, BW & 2018 & Journal & Information and Software Technology\\ \hline 

101 & SP-J48: a novel optimization and machine-learning-based approach for solving complex problems: special application in software engineering for detecting code smells & Amandeep Kaur, Sushma Jain, Shivani Goel & 2019 & Journal & Neural Computing and Applications \\ \hline 

102 & Visualizing code bad smells & Hammad, M., Alsofriya, S. & 2019 & Journal & International Journal of Advanced Computer Science and Applications\\ \hline 
\end{longtable}
%\end{table}
}
\pagebreak

\subsection{\large{Quality assessment}}
\label{AppendixC}

\footnotesize{
\begin{longtable}{|c|c|c|c|c|c|c|c|c|c|}
\hline
 Study & QC1& QC2& QC3& QC4& QC5& QC6& QC7& QC8 & Total\\
 & Venue Quality & Data Collected & Findings & Recognized Relevance & Validation & Replication & Evaluation & Visualization &  \\
\hline
S1 & 1 & 1 & 1 & 1 & 0 & 0 & 1 & 1 & 6 \\
S2 & 1 & 1 & 1 & 1 & 0 & 0 & 0 & 1 & 5 \\
S3 & 1 & 1 & 1 & 1 & 1 & 0 & 1 & 0 & 6 \\
S4 & 1 & 0 & 1 & 1 & 0 & 0 & 1 & 0 & 4 \\
S5 & 1 & 1 & 1 & 1 & 0 & 0 & 0 & 0 & 4 \\
S6 & 1 & 0 & 1 & 1 & 1 & 0 & 1 & 0 & 5 \\
S7 & 1 & 0 & 1 & 1 & 0 & 0 & 1 & 1 & 5 \\
S8 & 1 & 1 & 1 & 1 & 1 & 0 & 1 & 0 & 6 \\
S9 & 1 & 1 & 1 & 1 & 0 & 0 & 0 & 1 & 5 \\
S10 & 1 & 1 & 1 & 1 & 0 & 0 & 1 & 0 & 5 \\
S11 & 0 & 1 & 1 & 1 & 0 & 0 & 0 & 0 & 3 \\
S12 & 0 & 1 & 1 & 0 & 0 & 0 & 1 & 1 & 4 \\
S13 & 0 & 1 & 1 & 1 & 0 & 0 & 0 & 1 & 4 \\
S14 & 1 & 1 & 1 & 1 & 1 & 0 & 1 & 0 & 6 \\
S15 & 1 & 1 & 1 & 1 & 0 & 0 & 1 & 0 & 5 \\
S16 & 0 & 0 & 1 & 0 & 0 & 0 & 1 & 1 & 3 \\
S17 & 1 & 1 & 1 & 1 & 1 & 0 & 1 & 0 & 6 \\
S18 & 1 & 1 & 1 & 1 & 0 & 0 & 1 & 1 & 6 \\
S19 & 0 & 1 & 1 & 1 & 0 & 0 & 0 & 0 & 3 \\
S20 & 0 & 1 & 1 & 1 & 0 & 0 & 0 & 1 & 4 \\
S21 & 1 & 1 & 1 & 1 & 1 & 1 & 1 & 0 & 7 \\
S22 & 1 & 0 & 1 & 1 & 0 & 0 & 0 & 0 & 3 \\
S23 & 1 & 0 & 1 & 1 & 0 & 0 & 0 & 1 & 4 \\
S24 & 1 & 0 & 1 & 1 & 1 & 0 & 1 & 0 & 5 \\
S25 & 0 & 1 & 1 & 1 & 0 & 0 & 1 & 1 & 5 \\
S26 & 1 & 0 & 1 & 1 & 0 & 0 & 1 & 0 & 4 \\
S27 & 1 & 1 & 1 & 1 & 1 & 0 & 1 & 0 & 6 \\
S28 & 1 & 1 & 1 & 1 & 1 & 0 & 1 & 0 & 6 \\
S29 & 1 & 1 & 1 & 1 & 0 & 0 & 1 & 0 & 5 \\
S30 & 1 & 1 & 1 & 1 & 1 & 0 & 1 & 0 & 6 \\
S31 & 1 & 1 & 1 & 1 & 1 & 1 & 1 & 0 & 7 \\
S32 & 0 & 0 & 1 & 0 & 0 & 0 & 0 & 0 & 1 \\
S33 & 1 & 1 & 1 & 1 & 1 & 0 & 1 & 0 & 6 \\
S34 & 0 & 1 & 1 & 1 & 0 & 0 & 1 & 0 & 4 \\
S35 & 1 & 1 & 1 & 1 & 1 & 0 & 1 & 0 & 6 \\
S36 & 1 & 0 & 1 & 1 & 0 & 0 & 0 & 0 & 3 \\
S37 & 1 & 1 & 1 & 1 & 1 & 0 & 1 & 0 & 6 \\
S38 & 1 & 1 & 1 & 0 & 0 & 0 & 1 & 0 & 4 \\
S39 & 1 & 0 & 1 & 1 & 1 & 0 & 1 & 0 & 5 \\
S40 & 1 & 1 & 1 & 1 & 1 & 0 & 1 & 0 & 6 \\
S41 & 1 & 1 & 1 & 1 & 0 & 0 & 1 & 0 & 5 \\
S42 & 1 & 1 & 1 & 1 & 0 & 0 & 1 & 0 & 5 \\
S43 & 1 & 1 & 1 & 1 & 0 & 0 & 0 & 0 & 4 \\
S44 & 1 & 1 & 1 & 1 & 1 & 1 & 1 & 0 & 7 \\
S45 & 1 & 0 & 1 & 1 & 0 & 0 & 1 & 1 & 5 \\
S46 & 0 & 1 & 1 & 1 & 0 & 0 & 0 & 1 & 4 \\
S47 & 0 & 1 & 1 & 1 & 1 & 0 & 1 & 0 & 5 \\
S48 & 1 & 1 & 1 & 1 & 1 & 0 & 1 & 0 & 6 \\
S49 & 1 & 1 & 1 & 1 & 1 & 0 & 1 & 0 & 6 \\
S50 & 0 & 1 & 1 & 1 & 1 & 0 & 1 & 0 & 5 \\
S51 & 1 & 1 & 1 & 1 & 1 & 0 & 1 & 0 & 6 \\
S52 & 1 & 1 & 1 & 1 & 0 & 0 & 1 & 1 & 6 \\
S53 & 0 & 1 & 1 & 1 & 1 & 0 & 1 & 0 & 5 \\
S54 & 1 & 1 & 1 & 1 & 1 & 0 & 1 & 0 & 6 \\
S55 & 1 & 1 & 1 & 1 & 1 & 0 & 1 & 0 & 6 \\
S56 & 1 & 1 & 1 & 1 & 1 & 0 & 1 & 0 & 6 \\
S57 & 0 & 0 & 1 & 1 & 0 & 0 & 0 & 0 & 2 \\
S58 & 1 & 1 & 1 & 1 & 1 & 1 & 1 & 0 & 7 \\
S59 & 0 & 1 & 1 & 1 & 0 & 0 & 1 & 0 & 4 \\
S60 & 1 & 1 & 1 & 1 & 1 & 0 & 1 & 0 & 6 \\
S61 & 1 & 1 & 1 & 1 & 1 & 0 & 0 & 0 & 5 \\
S62 & 1 & 1 & 1 & 1 & 1 & 0 & 1 & 0 & 6 \\
S63 & 1 & 1 & 1 & 1 & 1 & 1 & 1 & 0 & 7 \\
S64 & 0 & 1 & 1 & 1 & 0 & 0 & 1 & 1 & 5 \\
S65 & 1 & 1 & 1 & 1 & 1 & 0 & 1 & 0 & 6 \\
S66 & 1 & 1 & 1 & 1 & 1 & 1 & 1 & 0 & 7 \\
S67 & 0 & 1 & 1 & 1 & 1 & 0 & 1 & 0 & 5 \\
S68 & 0 & 1 & 1 & 0 & 1 & 0 & 1 & 0 & 4 \\
S69 & 1 & 1 & 1 & 1 & 1 & 0 & 1 & 0 & 6 \\
S70 & 0 & 1 & 1 & 1 & 0 & 0 & 1 & 0 & 4 \\
S71 & 0 & 0 & 1 & 1 & 1 & 0 & 0 & 0 & 3 \\
S72 & 0 & 1 & 1 & 1 & 1 & 0 & 1 & 0 & 5 \\
S73 & 0 & 1 & 1 & 0 & 0 & 0 & 1 & 0 & 3 \\
S74 & 1 & 1 & 1 & 0 & 1 & 0 & 1 & 0 & 5 \\
S75 & 0 & 1 & 1 & 1 & 1 & 0 & 1 & 0 & 5 \\
S76 & 1 & 1 & 1 & 1 & 1 & 0 & 1 & 0 & 6 \\
S77 & 0 & 0 & 1 & 0 & 0 & 0 & 0 & 1 & 2 \\
S78 & 1 & 1 & 1 & 1 & 1 & 0 & 1 & 0 & 6 \\
S79 & 0 & 1 & 1 & 1 & 0 & 1 & 1 & 0 & 5 \\
S80 & 0 & 1 & 1 & 1 & 1 & 0 & 1 & 0 & 5 \\
S81 & 0 & 1 & 1 & 1 & 1 & 0 & 1 & 0 & 5 \\
S82 & 1 & 1 & 1 & 0 & 0 & 0 & 0 & 0 & 3 \\
S83 & 0 & 0 & 1 & 0 & 0 & 0 & 1 & 0 & 2 \\
S84 & 0 & 1 & 1 & 1 & 1 & 0 & 1 & 0 & 5 \\
S85 & 0 & 1 & 1 & 0 & 0 & 0 & 1 & 0 & 3 \\
S86 & 0 & 1 & 1 & 0 & 0 & 0 & 0 & 0 & 2 \\
S87 & 0 & 0 & 1 & 0 & 0 & 0 & 0 & 0 & 1 \\
S88 & 1 & 1 & 1 & 0 & 1 & 0 & 1 & 0 & 5 \\
S89 & 1 & 1 & 1 & 1 & 1 & 0 & 1 & 0 & 6 \\
S90 & 0 & 1 & 1 & 0 & 0 & 0 & 1 & 0 & 3 \\
S91 & 0 & 0 & 1 & 0 & 0 & 0 & 0 & 0 & 1 \\
S92 & 1 & 1 & 1 & 1 & 1 & 0 & 1 & 0 & 6 \\
S93 & 1 & 1 & 1 & 1 & 0 & 0 & 0 & 1 & 5 \\
S94 & 0 & 1 & 1 & 1 & 1 & 0 & 1 & 0 & 5 \\
S95 & 0 & 1 & 1 & 0 & 1 & 0 & 0 & 0 & 3 \\
S96 & 1 & 1 & 1 & 1 & 1 & 0 & 1 & 0 & 6 \\
S97 & 0 & 1 & 1 & 1 & 1 & 0 & 1 & 0 & 5 \\
S98 & 0 & 1 & 1 & 1 & 0 & 0 & 1 & 0 & 4 \\
S99 & 1 & 1 & 1 & 1 & 1 & 0 & 0 & 1 & 6 \\
S100 & 1 & 1 & 1 & 1 & 0 & 1 & 1 & 0 & 6 \\
S101 & 1 & 1 & 1 & 1 & 1 & 0 & 1 & 0 & 6 \\
S102 & 0 & 0 & 1 & 1 & 0 & 0 & 1 & 0 & 3 \\ \hline 
Total & 63 & 83 & 102 & 85 & 54 & 8 & 78 & 18 & 491\\
 \hline 
\end{longtable}
}
\pagebreak

\subsection{\large{Description of code smells detected in the studies}}
\label{AppendixD}

\begin{longtable}{|p{5cm}|p{10cm}|c|}
\hline
\textbf {Code smell} & \textbf{Description} & \textbf{Reference}\\
\hline
Alternative Classes with Different Interface & One class supports different classes, but their interface is different  & \cite{Fowler1999,LACERDA2020} 
\\ \hline
AntiSingleton                               & A class that provides mutable class variables, which consequently could be used as global variables  & \cite{Khomh2012}
\\ \hline
God Class (Large Class or Blob)              & Class that has many responsibilities and therefore contains many variables and methods. The same Single Responsibility Principle (SRP) also applies in this case & \cite{Fowler1999,LACERDA2020}
\\ \hline
Brain Class                                  & Class tend to be complex and centralize the functionality of the system. It is therefore assumed that they are difficult to understand and maintain. However, contrary to God Classes, Brain Classes do not use much data from foreign  classes and are slightly more cohesive   & \cite{Lanza2006,Olbrich2010}
\\ \hline
Brain Method                                 & Often a method starts out as a “normal” method but then more and more functionality is added to it until it gets out of control, becoming hard to maintain or understand. Brain Methods tend to centralize the functionality of a class  & \cite{Lanza2006}
\\ \hline
Careless Cleanup                             & The exception resource can be interrupted by another exception  &  \cite{Gupta2018}
\\ \hline
Class Data Should Be Private                 & A class that exposes its fields, thus violating the principle of encapsulation  &  \cite{Khomh2012}
\\ \hline
Closure Smells                               & In JavaScript, it is possible to declare nested functions, called closures. Closures make it possible to emulate object oriented notions such as \textit{public, private}, and \textit{privileged} members. Inner functions have access to the parameters and variables — except for \textit{this} and \textit{argument} variables — of the functions they are nested in, even after the outer function has returned. Four smells related to the concept of function closures (long scope chaining, closures in loops, variable name conflict in closures, accessing the this reference in closures) & \cite{Fard2013}
\\ \hline
Code clone/Duplicated code                   & Consists of equal or very similar passages in different fragments of the same code base  & \cite{Fowler1999,LACERDA2020} 
\\ \hline
Comments                                & It cannot be considered a smell by definition but should be used with care as they are generally not required. Whenever it is necessary to insert a comment, it is worth checking if the code cannot be more expressive  & \cite{Fowler1999,LACERDA2020}
\\ \hline
Complex Class                                & A class that has (at least) one large and complex method, in terms of cyclomatic complexity and LOCs  & \cite{Khomh2012}
\\ \hline
Complex Container Comprehension              & A container comprehension (including list comprehension, set comprehension, dictionary comprehension, and generator expression) that is too complex  & \cite{Chen2016} 
\\ \hline
Complex List Comprehension                   & A list comprehension that is too complex. List comprehensions in Python provide a concise and efficient way to create new lists, especially where the value of each element is dependent on each member of another iterable or sequence, or the elements satisfy a certain condition. However, when list comprehensions contain complex expressions, they are no longer clear. Apparently, it is hard to analyze control flows of complex list comprehensions  & \cite{Chen2016} \\ \hline
Coupling between JavaScript, HTML, and CSS   &  In web applications, HTML is meant for presenting content and structure, CSS for styling, and JavaScript for functional behaviour. Keeping these three entities separate is a well-known programming practice, known as separation of concerns. Unfortunately, web developers often mix JavaScript code with markup and styling code, which adversely influences program comprehension, maintenance and debugging efforts in web applications &  \cite{Fard2013}
\\ \hline
Data class                                   & The class that serves only as a container of data, without any behavior. Generally, other classes are responsible for manipulating their data, which is a case of Feature Envy & \cite{Fowler1999,LACERDA2020}
\\ \hline
Data Clump                                   & Data structures that always appear together, and when one of the items is not present, the whole set loses its meaning  & \cite{Fowler1999,LACERDA2020}
\\ \hline
Dead Code                                    & Characterized by a variable, attribute, or code fragment that is not used anywhere. It is usually a result of a code change with improper cleaning  & \cite{Wake2003,LACERDA2020}
\\ \hline
Delegator                                    & Overuse of delegation or misuse of inheritance  & \cite{Kreimer2005} 
\\ \hline
Dispersed Coupling                           & Refers to a method which is tied to many operations dispersed among many classes throughout the system  & \cite{Lanza2006}
\\ \hline
Divergent Change                             & A single class needs to be changed for many reasons. This is a clear indication that it is not sufficiently cohesive and must be divided  & \cite{Fowler1999,LACERDA2020}
\\ \hline
Dummy Handler                                & Dummy handler is only used for viewing the exception but it will not handle the exception  &  \cite{Gupta2018} 
\\ \hline
Empty Catch Block                            & When the catch block is left blank in the catch statement  & \cite{Gupta2018}
\\ \hline
Exception thrown in the finally block        & How to handle the exception thrown inside the finally block of another try catch statement  &  \cite{Gupta2018} 
\\ \hline
Excessive Global Variables                   & Global variables are accessible from anywhere in JavaScript code, even when defined in different files loaded on the same page. As such, naming conflicts between global variables in different JavaScript source files is common, which affects program dependability and correctness. The higher the number of global variables in the code, the more dependent existing modules are likely to be; and dependency increases errorproneness, and maintainability efforts  & \cite{Fard2013} 
\\ \hline
Feature Envy                                 & When a method is more interested in members of other classes than its own, is a clear sign that it is in the wrong class & \cite{Fowler1999,LACERDA2020}
\\ \hline
Functional Decomposition                     & A procedural code in a technology that implements the OO paradigm (usually the main function that calls many others), caused by the previous expertise of the developers in a procedural language and little experience in OO & \cite{Brown1998,LACERDA2020}\\ 
\hline
God Package                                  & A package that is too large. That knows too much or does too much  &   \cite{Martin2002}                           
\\ \hline
Inappropriate Intimacy                       & A case where two classes are known too, characterizing a high level of coupling  & \cite{Fowler1999,LACERDA2020} 
\\ \hline
Incomplete Library Class                     & The software uses a library that is not complete, and therefore extensions to that library are required  & \cite{Fowler1999,LACERDA2020}  
\\ \hline
Instanceof                                   & In Java, the instanceof operator is used to check that an object is an instance of a given class or implements a certain interface. These are considered CS aspects because a concentration of instanceof operators in the same block of code may indicate a place where the introduction of an inheritance hierarchy or the use of method overloading might be a better solution  & \cite{Emden2002} 
\\ \hline
Intensive Coupling                           & Refers to a method that is tied to many other operations located in only a few classes within the system.  & \cite{Lanza2006}
\\ \hline
Introduce null object                        & Repeated null checking conditions are added into the code to prevent the null pointer exception problem. By doing so, the duplications of null checking conditions could have been placed in different locations of the software system  & \cite{Sirikul2016}  
\\ \hline
Large object                                 & An object with too many responsibilities. An object that is doing too much should be refactored. Large objects may be restructured or broken into smaller objects & \cite{Fard2013}
\\ \hline
Lazy Class                                   & 	Classes that do not have sufficient responsibilities and therefore should not exist  & \cite{Fowler1999,LACERDA2020}
\\ \hline
Lazy object                                  & An object that does too little. An object that is not doing enough work should be refactored. Lazy objects maybe collapsed or combined into other classes & \cite{Fard2013}
\\ \hline
Long Base Class List                         & A class definition with too many base classes. Python supports a limited form of multiple inheritance. If an attribute in Python is not found in the derived class during execution, it is searched recursively in the base classes declared in the base class list in sequence. Too long base class list will limit the speed of interpretive execution  & \cite{Chen2016}
\\ \hline
Long Element Chain                           & An expression that is accessing an object through a long chain of elements by the bracket operator. Long Element Chain is directly caused by nested arrays. It is unreadable especially when a deep level of array traversing is taking place  & \cite{Chen2016}  
\\ \hline
Long Lambda Function                         & A lambda function that is overly long, in term of the number of its characters. Lambda is a powerful construct that allows the creation of anonymous functions at runtime. A lambda function can only contain one single expression. If lambda is overly long (i.e. it contains too complex operations), it turns out to be unreadable and loses its benefits. In order to avoid too many one-expression long functions in Python programs, lambda should only be used in packaging special or non-reusable code, otherwise explicit def statements can always take the place of them  & \cite{Chen2016}  
\\ \hline
Long Message Chain                           &
An expression that is accessing an object through a long chain of attributes or methods by the dot operator  & \cite{Chen2016} 
\\ \hline
Long Method                                  & Very large method/function and, therefore, difficult to understand, extend and modify. It is very likely that this method has too many responsibilities, hurting one of the principles of a good OO design (SRP: Single Responsibility Principle & \cite{Fowler1999,LACERDA2020}
\\ \hline
Long Parameter List                          & Extensive parameter list, which makes it difficult to understand and is usually an indication that the method has too many responsibilities. This smell has a strong relationship with Long Method & \cite{Fowler1999,LACERDA2020} 
\\ \hline
Long Scope Chaining                          & A method or a function that is multiply-nested  & \cite{Chen2016} 
\\ \hline
Long Ternary Conditional Expression          & A ternary conditional expression that is overly long. The ternary operator defines a new conditional expression in Python, with the value of the expression being X or Y based on the truth value of C in the form of "X if C else Y". However, it is rather long when it involves other constructs such as lambda functions. Saving a few characters in it would be difficult when it contains several long variable names. As a result, though the ternary conditional expression is a concise way of the conditional expression to help avoid ugly, a long one is unreadable. The solution to refactoring is to transform it into a traditional conditional expression  & \cite{Chen2016}
\\ \hline
Message Chain                                & One object accesses another, to then access another object belonging to this second, and so on, causing a high coupling between classes  & \cite{Fowler1999,LACERDA2020} 
\\ \hline
Method call sequences                        & The interplay of multiple methods, though—in particular, whether a specific sequence of method calls is allowed or not—is neither specified nor checked at compile time. Consequently, illegal call sequences may still loom in the code even though all tests pass.  &  \cite{Wasylkowski2007}
\\ \hline
Middle Man                                   & Identified how much a class has almost no logic, as it delegates almost everything to another class. The problem with this CS is that whenever you need to create new methods or to modify the old ones, you also have to add or modify the delegating method    & \cite{Fowler1999,LACERDA2020}
\\ \hline
Misplaced Class                              & Suggests a class that is in a package that contains other classes not related to it. The obvious way to remove such a smell is to apply a Move Class refactoring able to place the class in a more related package.  & \cite{Palomba2016}
\\ \hline
Missing method calls                         & Overlook certain important method calls that are required at particular places in code  &\cite{Monperrus2010} 
\\ \hline
Multiply-Nested Container                    & A container (including set, list, tuple, dict) that is multiply-nested. It directly produces expressions accessing an object through a long chain of indexed elements. It is unreadable especially when there is a deep level of array traversing  & \cite{Chen2016}
\\ \hline
Nested Callback                              & A callback is a function passed as an argument to another (parent) function. Callbacks are executed after the parent function has completed its execution. Callback functions are typically used in asynchronous calls such as timeouts and XMLHttpRequests (XHRs). Using excessive callbacks, however, can result in hard to read and maintain code due to their nested anonymous (and usually asynchronous) nature  &\cite{Fard2013}
\\ \hline
Nested Try Statements                        & When one or more try statements are contained in the try statement  & \cite{Gupta2018}
\\ \hline
Null checking in a string comparison problem & Null checking conditions are usually found in string comparison, particularly in an if statement. This form of defensive programming can be employed to prevent the null pointer exception error. The same null checking statement is repeatedly appeared when the same String object is compared, resulting in a marvellous number of duplicated null checking conditions. The null checking conditions could be eliminated by calling the SomeString.equals(value) method instead  & \cite{Sirikul2016} 
\\ \hline
Parallel Inheritance                     & Existence of two hierarchies of classes that are fully connected, that is, when adding a subclass in one of the hierarchies, it is required that a similar subclass be created in the other  & \cite{Fowler1999,LACERDA2020} 
\\ \hline
Primitive Obsession                          & It represents the situation where primitive types are used in place of light classes  & \cite{Fowler1999,LACERDA2020} 
\\ \hline
Promiscuous Package                          & A package can be considered as promiscuous if it contains classes implementing too many features, making it too hard to understand and maintain  &\cite{Palomba2016} 
\\ \hline
Refused Bequest                              & 	It indicates that a subclass does not use inherited data or behaviors & \cite{Fowler1999,LACERDA2020}  
\\ \hline
Shotgun Surgery                              & Opposite to Divergent Change, because when it happens a modification, several different classes have to be changed & \cite{Fowler1999,LACERDA2020}  
\\ \hline
Spaghetti Code                               & Use of classes without structures, long methods without parameters, use of global variables, in addition to not exploiting and preventing the application of OO principles such as inheritance and polymorphism & \cite{Brown1998,LACERDA2020}
\\ \hline
Speculative Generality                       & Code snippets are designed to support future software behavior that is not yet required  & \cite{Fowler1999,LACERDA2020}
\\ \hline
Swiss Army Knife                             & Exposes the high complexity to meet the predictable needs of a part of the system (usually utility classes with many responsibilities) & \cite{Brown1998,LACERDA2020}
\\ \hline
Switch Statement                             & It is not necessarily smells by definition, but when they are widely used, they are usually a sign of problems, especially when used to identify the behavior of an object based on its type  & \cite{Fowler1999,LACERDA2020} 
\\ \hline
Temporary Field                              & Member-only used in specific situations, and that outside of it has no meaning  & \cite{Fowler1999,LACERDA2020}
\\ \hline
Tradition Breaker                            & This design disharmony strategy takes its name from the principle that the interface of a class (i.e., the services that it provides to the rest of the system) should increase in an evolutionary fashion. This means that a derived class should not break the inherited “tradition” and provide a large set of services which are unrelated to those provided by its base class.  & \cite{Lanza2006}
\\ \hline
Type Checking                                & Type-checking code is introduced in order to select a variation of an algorithm that should be executed, depending on the value of an attribute. Mainly it manifests itself as complicated conditional statements that make the code difficult to understand and maintain  & \cite{Tsantalis2008}
\\ \hline
Typecast                                     & Typecasts are used to explicitly convert an object from one class type into another. Many people consider typecasts to be problematic since it is possible to write illegal casting instructions in the source code which cannot be detected during compilation but result in runtime errors  & \cite{Emden2002} 
\\ \hline
Unprotected Main                             & Outer exception will not be handled in the main program; it can only be handled in a subprogram or a function  & \cite{Gupta2018}  
\\ \hline
Useless Exception Handling                   & A try...except statement that does little. A try statement in Python can have more than one except clause to specify handlers for different selected exceptions. There are two kinds of writing styles which make exception handling useless. First, it has only one except clause and it catches a too general exception such as Exception and StandardError or even can catch all exceptions. The damage of this writing style is hiding the true exception in the try clause. Second, all exception clauses in the statement are empty, which means whatever exceptions in the try clause will not be responded  & \cite{Chen2016} 
\\ \hline
Wide Subsystem Interface                     & A Subsystem Interface consists of classes that are accessible from outside the package they belong to. The flaw refers to the situation where this interface is very wide, which causes a very tight coupling between the package and the rest of the system  & \cite{Yamashita2012} 
\\ \hline
\end{longtable}

\pagebreak

\subsection{\large{Frequencies of code smells detected in the studies}}
\label{AppendixE}

\begin{longtable}{|p{5cm}|c|c|l|}
\hline
\textbf {Code smell} & \textbf{Nº of studies} & \textbf{\% Studies} & \textbf{programming language}\\
\hline
God Class (Large Class or Blob)              & 43 & 51.8\% & Java, C/C++ , C\#, Python             \\ \hline
Feature Envy                                 & 28 & 33.7\% & Java, C/C++ , C\#                     \\ \hline
Long Method                                  & 22 & 26.5\% & Java, C/C++ , C\#, Python, JavaScript \\ \hline
Data class                                   & 18 & 21.7\% & Java, C/C++ , C\#                     \\ \hline
Functional Decomposition                     & 17 & 20.5\% & Java                                  \\ \hline
Spaghetti Code                               & 17 & 20.5\% & Java                                  \\ \hline
Long Parameter List                          & 12 & 14.5\% & Java, C/C++ , C\#, Python, JavaScript \\ \hline
Swiss Army Knife                             & 11 & 13.3\% & Java                                  \\ \hline
Refused Bequest                              & 10 & 12.0\% & Java, C/C++ , C\#, JavaScript         \\ \hline
Shotgun Surgery                              & 10 & 12.0\% & Java, C++ , C\#                       \\ \hline
Code clone/Duplicated code                   & 9  & 10.8\%  & Java, C/C++ , C\#                     \\ \hline
Lazy Class                                   & 8  & 9.6\%  & Java, C++ , C\#                       \\ \hline
Divergent Change                             & 7  & 8.4\%  & Java, C\#                             \\ \hline
Dead Code                                    & 4  & 4.8\%  & Java, C++ , C\#                       \\ \hline
Switch Statement                             & 4  & 4.8\%  & Java, C\#, JavaScript                 \\ \hline
Brain Class                                  & 3  & 3.6\%  & Java, C++                             \\ \hline
Data Clump                                   & 3  & 3.6\%  & Java, C/C++ , C\#                     \\ \hline
Long Message Chain                           & 3  & 3.6\%  & JavaScript, Python                  \\ \hline
Misplaced Class                             & 3  & 3.6\%  & Java, C++                            \\ \hline
Parallel Inheritance                         & 3  & 3.6\%  & Java, C\#                             \\ \hline
Primitive Obsession                          & 3  & 3.6\%  & Java, C/C++ , C\#                     \\ \hline
Speculative Generality                       & 3  & 3.6\%  & Java, C\#                             \\ \hline
Temporary Field                              & 3  & 3.6\%  & Java, C\#                             \\ \hline
Dispersed Coupling                           & 2  & 2.4\%  & Java, C++                             \\ \hline
Empty Catch Block                            & 2  & 2.4\%  & Java, JavaScript                      \\ \hline
Excessive Global Variables                   & 2  & 2.4\%  & JavaScript                            \\ \hline
Intensive Coupling                           & 2  & 2.4\%  & Java, C++                             \\ \hline
Large object                                 & 2  & 2.4\%  & JavaScript                            \\ \hline
Lazy object                                  & 2  & 2.4\%  & JavaScript                            \\ \hline
Long Base Class List                         & 2  & 2.4\%  & Python                                \\ \hline
Long Lambda Function                         & 2  & 2.4\%  & Python                                \\ \hline
Long Scope Chaining                          & 2  & 2.4\%  & Python                                \\ \hline
Long Ternary Conditional Expression          & 2  & 2.4\%  & Python                                \\ \hline
Message Chain                                & 2  & 2.4\%  & Java, C/C++ , C\#                     \\ \hline
Middle Man                                   & 2  & 2.4\%  & Java, C/C++ , C\#                     \\ \hline
Missing method calls                         & 2  & 2.4\%  & Java                                  \\ \hline
Alternative Classes with Different Interface & 1  & 1.2\%  & Java, C\#                             \\ \hline
AntiSingleton                               & 1  & 1.2\%  & Java                                  \\ \hline
Brain Method                                 & 1  & 1.2\%  & Java, C++                             \\ \hline
Careless Cleanup                             & 1  & 1.2\%  & Java                                  \\ \hline
Class Data Should Be Private                 & 1  & 1.2\%  & Java                                  \\ \hline
Closure Smells                               & 1  & 1.2\%  & JavaScript                            \\ \hline
Comments                                & 1  & 1.2\%  & Java, C\#                             \\ \hline
Complex Class                                & 1  & 1.2\%  & Java                                  \\ \hline
Complex Container Comprehension              & 1  & 1.2\%  & Python                                \\ \hline
Complex List Comprehension                   & 1  & 1.2\%  & Python                                \\ \hline
Coupling between JavaScript, HTML, and CSS   & 1  & 1.2\%  & JavaScript                            \\ \hline
Delegator                                    & 1  & 1.2\%  & Java                                  \\ \hline
Dummy Handler                                & 1  & 1.2\%  & Java                                  \\ \hline
Exception thrown in the finally block        & 1  & 1.2\%  & Java                                  \\ \hline
God Package                                  & 1  & 1.2\%  & Java, C++                             \\ \hline
Inappropriate Intimacy                       & 1  & 1.2\%  & Java, C\#                             \\ \hline
Incomplete Library Class                     & 1  & 1.2\%  & Java, C\#                             \\ \hline
Instanceof                                   & 1  & 1.2\%  & Java                                  \\ \hline
Introduce null object                        & 1  & 1.2\%  & Java                                  \\ \hline
Long Element Chain                           & 1  & 1.2\%  & Python                                \\ \hline
Method call sequences                        & 1  & 1.2\%  & Java                                  \\ \hline
Multiply-Nested Container                    & 1  & 1.2\%  & Python                                \\ \hline
Nested Callback                              & 1  & 1.2\%  & JavaScript                            \\ \hline
Nested Try Statements                        & 1  & 1.2\%  & Java                                  \\ \hline
Null checking in a string comparison problem & 1  & 1.2\%  & Java                                  \\ \hline
Promiscuous Package                          & 1  & 1.2\%  & Java                                  \\ \hline
Tradition Breaker                            & 1  & 1.2\%  & Java, C++                             \\ \hline
Type Checking                                & 1  & 1.2\%  & Java                                  \\ \hline
Typecast                                     & 1  & 1.2\%  & Java                                  \\ \hline
Unprotected Main                             & 1  & 1.2\%  & Java                                  \\ \hline
Useless Exception Handling                   & 1  & 1.2\%  & Python                                \\ \hline
Wide Subsystem Interface                     & 1  & 1.2\%  & Java, C++                             \\ \hline
\end{longtable}

\end{document}